# Materials and possible mechanisms of extremely large magnetoresistance: A review

Rui Niu[1,2] and W. K. Zhu[1]

[1] High Magnetic Field Laboratory, Chinese Academy of Sciences, Hefei 230031, China

[2] University of Science and Technology of China, Hefei 230026, China

E-mail: wkzhu@hmfl.ac.cn

## Abstract

Magnetoresistance (MR) is a characteristic that the resistance of a substance changes with the external magnetic field, reflecting various physical origins and microstructures of the substance. A large MR, namely a huge response to a low external field, has always been a useful functional feature in industrial technology and a core goal pursued by physicists and materials scientists. Conventional large MR materials are mainly manganites, whose colossal MR (CMR) can be as high as -90%. The dominant mechanism is attributed to spin configuration aligned by the external field, which reduces magnetic scattering and thus resistance. In recent years, some new systems have shown an extremely large unsaturated MR (XMR). Unlike ordinary metals, the positive MR of these systems can reach $10^3$-$10^8$% and is persistent under super high magnetic fields. The XMR materials are mainly metals or semimetals, distributed in high-mobility topological or non-topological systems, and some are magnetic, which suggests a wide range of application scenarios. Various mechanisms have been proposed for the potential physical origin of XMR, including electron-hole compensation, steep band, ultrahigh mobility, high residual resistance ratio, topological fermions, etc. It turns out that some mechanisms play a leading role in certain systems, while more are far from clearly defined. In addition, the researches on XMR are largely overlapped or closely correlated with other recently rising physics and materials researches, such as topological matters and two-dimensional (2D) materials, which makes elucidating the mechanism of XMR even more important. Moreover, the disclosed novel properties will lay a broad and solid foundation for the design and development of functional devices. In this review, we will discuss several aspects in the following order: (I) Introduction, (II) XMR materials and classification, (III) Proposed mechanisms for XMR, (IV) Correlation with other systems (featured), and (V) Conclusions and outlook.

## 1. Introduction

Magnetoresistance (MR) is the change in electrical resistivity of a material exposed to an external magnetic field. MR ratio, an index to evaluate MR, is defined as $\mathrm{MR} = \frac{\Delta\rho}{\rho(0)} = \frac{\rho(B)-\rho(0)}{\rho(0)}$, where $\rho(B)$ and $\rho(0)$ are the resistivity measured with and without magnetic field *B*, respectively. Like most other parameters of crystals, MR is anisotropic. When the external field is parallel to the current (*I*), MR is called longitudinal MR (LMR). When *B* is normal to *I*, MR is called transverse MR. Hereinafter, if not specified, the MR involved with a value refers to the MR ratio of transverse MR. Since William Thomson discovered the MR effect in 1856, the system has been extensively studied and applied in a wide range of fields such as magnetic storage, sensing, control and computation. Especially for magnetic memory, with the increasing demand for working under complex and harsh conditions, people are committed to finding two dimensional (2D) materials with greater MR and better performance to meet the requirements for access speed, capacity, sensitivity, stability and miniaturization of devices.

In most metals, resistivity increases as magnetic field increases, that is, MR is positive, while in some transition metal alloys and ferromagnets, MR is negative [1]. According to the magnitude and mechanism of MR, MR phenomena can be divided into ordinary MR (OMR), giant MR (GMR), colossal MR (CMR) and tunneling MR (TMR). Unlike these conventional MR phenomena, an extremely large MR (XMR) has been discovered and aroused tremendous interest in recent years. XMR means that the MR of certain metals or semimetals increases rapidly at low temperatures, as high as $10^3$-$10^8$%, and more importantly, the MR is not saturated under super high magnetic fields. For instance, no signature of saturation is detected for the XMR of $WTe_2$ under a magnetic field up to 60 T [2]. In this section, we will first introduce the history of MR research, then focus on a brief introduction to XMR materials and proposed mechanisms that will be detailed in the following sections, and then describe the relationship between XMR and 2D materials and topological materials.

For non-magnetic metals, the movement of carriers will be deflected or spiraled under the Lorentz force in the magnetic field, which will increase the probability of carrier scattering, thereby increasing the resistance. For general non-magnetic metals, such an OMR is very small, usually less than 5% [3]. At low temperatures and low fields, MR roughly follows the $B^2$ law. Unlike the positive OMR, GMR is a negative value (and usually greater than OMR) that occurs in magnetic materials. In 2007, Albert Fert and Peter Grünberg were awarded the Nobel Prize in physics for their discovery of GMR. Magnetic multilayer structure, consisting of alternately stacking ferromagnetic layer and non-ferromagnetic layer, can produce a GMR. When the magnetic moments of ferromagnetic layers are parallel to each other, the scattering is weak and the resistance is small; when the magnetic moments are antiparallel to each other, the scattering is strong and the resistance is large [4]. Since the external field tends to align the moments in a parallel manner, the GMR is negative. CMR represents the MR in manganites with a perovskite structure, which is usually much larger than GMR. The generation mechanism of CMR is different from that of GMR, and it is still controversial. It is currently believed that CMR may originate from double exchange interaction, Jahn-Teller distortion and phase separation. TMR refers to the phenomenon that the tunneling resistance of a sandwich structure, namely magnetic layer-insulating layer-magnetic layer, changes with the direction of magnetic moments of the magnetic layers. When the interlayer magnetic moments are parallel between layers, the probability of electrons tunneling through the insulating layer is high and the resistance is small; when the moments are antiparallel, the probability is low and the resistance is large. TMR is an important device prototype.

XMR has been observed in dozens of materials. A ready-to-use classification method is based on chemical composition. We divide the XMR materials that have been intensively studied in the past few years into elemental materials, binary and ternary compounds. Typical elemental XMR materials include Bi [5], α-As [6], α-Ga [7] and Sn [8]. As a semimetal, Bi has a very low carrier concentration, a small effective mass $m^*$ [9] and an impressively high carrier mobility [10]. These factors are considered to be related to XMR.

Several types of binary XMR materials are of particular interest, i.e., XP(As), XSb(Bi), transition metal dichalcogenides (TMDs), etc. XP and XAs represent four topological semimetals where X stands for Ta or Nb atom. These compounds lack space inversion symmetry but maintain time reversal symmetry. They are characterized to be type I Weyl semimetals [11-14]. The research interest in nonmagnetic monopnictides with formula XSb or XBi (X represents Y, Sc and lanthanides) focuses on the "resistance plateau" appearing at low temperatures, while the interest in magnetic monopnictides is to find magnetic topological semimetals that lack time reversal symmetry. Although LaSb is topologically trivial [15], its magnetic analogue, NdSb, is a potential candidate for magnetic topological semimetal, which is supported by the characterization of angle-resolved photoelectron spectroscopy (ARPES) [16] and the observation of negative LMR that may be caused by chiral anomaly [17]. TMDs are important layered compounds with a wide range of transport properties varying between different phases and chemical compositions. XMR commonly appears in tellurides that are semimetals or metals in nature, including $WTe_2$, $MoTe_2$, $PdTe_2$ and $PtTe_2$ [2, 18-20]. The layers of TMDs are connected by relatively weak van der Waals (vdW) force. Therefore, the crystals can be easily cleaved or peeled off by physical or chemical force. It is found that the XMR of TMDs is highly dependent on the thickness or number of layers. Other than TMDs, some compounds with formula $AB_2$ also have a significant XMR. For example, the XMR in $WP_2$ can be as high as $2 \times 10^8\%$ at 2.5 K and 63 T, which is attributed to the combination of electron-hole compensation, high mobility and extremely large residual resistance ratio (RRR) [21]. $ZrTe_5$ and $HfTe_5$ are another pair of binary compounds that simultaneously show semimetal transport and XMR characteristic, though their topological nature is still confusing. A topological transition can be easily achieved by applying an external field [22] or hydrostatic pressure [23].

Ternary materials such as half-Heusler alloys, $Co_3Sn_2S_2$ and ZrSiS/ZrSiSe have been intensively studied for their novel properties of magnetic topological semimetals or nodal-line semimetals. As a representative nodal-line semimetal, ZrSiSe not only has the common characteristics of Dirac semimetals, i.e., ultrahigh mobility and obvious quantum oscillations, but also has an outstanding XMR up to $10^8\%$ at 4.2 K and 60 T [24]. However, for GdPtBi and $Co_3Sn_2S_2$, the MR is only a small value ~ 200% [25, 26], which fails to reach the lower limit of XMR (set to $10^3$% in this review). Although they are both a magnetic topological semimetal that is supposed to favor a significant XMR, the XMR is absent. This may be due to the existence of magnetic ordering and magnetic scattering.

For XMR, various potential mechanisms have been proposed, including electron-hole compensation, steep band, ultrahigh mobility, high RRR, topological fermions, etc. The compensation mechanism refers to the balanced carrier density of two bands of charge carriers, i.e., electron-type and hole-type carriers, which is a highly simplified model. Although the electron-hole compensation is often considered to be the main mechanism of XMR, it has some limitations. The model fails to describe the complex cases of multiple bands with distinctly different effective masses and mobilities. In addition, XMR can still survive in certain situations that are far from compensation. This suggests that the compensation mechanism is unlikely to account for the (only) origin of XMR. High mobility is another factor that people often use to explain XMR. In the two-band model, MR is derived as $(\mu_{\mathrm{avg}}B)^2$ for the perfect compensation, i.e., the Lorentz law, where $\mu_{\mathrm{avg}}$ is the average mobility of carriers. High mobility is always a driving factor of XMR. As we know, mobility is inversely proportional to effective mass, and a small area of Fermi surface results in carriers of small effective mass. This explains why most XMR materials are semimetals that have small pockets and minor density of states (DOS) at the Fermi level.

XMR also relies on the crystal quality of samples which is usually represented by RRR. It is found that the correlation of $\mu_{\mathrm{avg}}$ with RRR appears roughly linear [27], which means that the scattering caused by impurities and defects will be suppressed in higher quality crystals. This is direct evidence for the positive impact of RRR on XMR and again proves the effect of mobility. Steep band refers to linear or nearly linear energy dispersion, regardless of whether there is band crossing or inversion. Since linear dispersion corresponds to massless carriers, the steep band means an ultrahigh mobility in favor of XMR. In real metals, normal parabolic dispersion is related to trivial electrons, and steep bands usually point to special or nontrivial electrons. The relationship between XMR and topological mechanism in topological materials is mainly reflected in two aspects. First, the electronic feature of topological materials is that there are band crossings in the momentum space and the energy bands are linear or nearly linear. Second, for topological systems, the spin and momentum of the carriers in the topological surface state are locked, and a conductive channel without backscattering is formed on the surface of the material, which improves the carrier mobility. In addition, there are some unusual mechanisms that can generate XMR or work together with one or more of the mechanisms listed above. Orbital character is a factor that cannot be ignored. For instance, the XMR of $MoAs_2$ is linked to the carrier motion on the Fermi surfaces with dominant open-orbit topology [28].

In most cases, compensation seems to be the primary driving force of MR. However, this does not mean that perfect compensation is necessary. For cases with poor compensation (sometimes far from compensation), XMR can still exist. The factor that comes in the second is high mobility which always promotes a large MR. We may realize that some of the mechanisms discussed above are more or less related to carrier mobility, including RRR, steep band and topological protection. Therefore, compensation and mobility are the first two factors to consider. In-depth research involves the structure of Fermi surfaces and the carrier properties that differ between materials.

It is found that XMR materials are largely overlapped or closely correlated with layered and 2D materials and topological matters, both of which are the mainstream of current condensed matter physics. Indeed, XMR was first proposed in the research of 2D materials and topological materials, and it has quickly grown into a large family and become a hot topic. This fact makes it even more important to clarify the mechanism of XMR. On the one hand, as a fundamental parameter, XMR reflects the basic features of transport and the key information of Fermi surface. XMR research can definitely help understand the origin of nontrivial properties in topological materials and 2D materials. On the other hand, the disclosed new properties of interdisciplinary fields will lay a solid and broad foundation for the design and development of functional devices.

Here 2D materials refer to layered materials that can be exfoliated into thin flakes or nanosheets (several layers) or even a single layer, rather than narrowly 2D monolayers. One advantage of 2D materials is the possibility to fabricate miniaturized devices. For 2D materials, carrier migration is limited to the 2D plane and suppressed between planes, which induces many unique transport properties. The electronic structure largely depends on the thickness or layers of the material, so does XMR. TMDs like $WTe_2$ and $MoTe_2$ are representative 2D XMR materials.

Many characteristics of topological materials are promotive to XMR, such as suppressed backscattering by topological protection, steep band, small effective mass and high mobility. The body state of topological insulators is an insulator, and the surface state is protected by topology and presents a metallic state. Electrons can be transported in two conductive channels on the surface of the material, and the direction of motion and the direction of spin are locked, i.e., spin momentum locking. Owning to this feature, the transport process of the surface state is not scattered. Another large group of topological materials is topological semimetals, which are characterized by symmetry protected band crossings at or near the Fermi level in the Brillouin zone. Compared with topological insulators, the XMR in topological semimetals is more prominent, due to the semimetal nature. According to the degeneracy and momentum space distribution of the nodal points, topological semimetals can be divided into three categories, i.e., Dirac semimetals, Weyl semimetals (type I and II) and nodal-line semimetals. For each category, we discuss one or two representative materials, namely $Cd_3As_2$, ZrSiS, TaAs, $WTe_2$ ($MoTe_2$) and $Co_3Sn_2S_2$. Except for $Co_3Sn_2S_2$, all semimetals exhibit a significant XMR. The relationship between XMR and topological properties is discussed, as well as the dependence on external factors such as temperature, sample shape and size.

To summarize this section, XMR is a fundamental, important and fruitful research direction in modern condensed matter physics, for both theoretical and experimental researchers. In this review, for indexing convenience, all XMR materials and their measurement conditions are tabulated and plotted. A table is also drawn to present the proposed mechanisms and research methods of XMR materials. Based on the comprehensive review and thorough discussion, a perspective outlook is provided as a concluding remark.

## 2. XMR materials and classification

In this section, we present XMR materials that have been intensively studied in the past few years, and divide them into elemental materials, binary and ternary compounds according to chemical composition. These compounds are further classified in terms of their chemical formula, crystal structure and/or other common physical properties, as listed in various subsections. Note that in each subsection, only representative materials are shown and discussed in detail. Other prominent systems are presented in the mechanism section to avoid repetitive discussions.

### 2.1 Elemental materials

Pronounced XMR has been observed in elemental materials, such as Bi [5], α-As [6] and α-Ga [7]. Table 1 lists the XMR and measurement conditions of elemental materials.

**Table 1.** XMR and measurement conditions of elemental materials.

| Material | Measurement conditions | MR (%) | Reference |
|---|---|---|---|
| Bi | 5 K, 5 T | $3.8 \times 10^5$ | [5] |
| α-As | 1.8 K, 9 T | $1.5 \times 10^7$ | [6] |
| α-Ga | 2 K，9 T | $1.66 \times 10^6$ | [7] |

Bulk Bi possesses a typical rhombohedral structure (*R*-3*m* space group, A7 structure), as illustrated in figure 1(a) [29]. Bi single crystal is a layered material. In the [111] direction of the rhombohedral structure, the two nearest-neighboring atoms are not at the same height and a layered structure consisting of diatomic layers is formed. As the covalent bonds between atoms are much stronger than the interlayer binding energy, Bi single crystal is easy to cleave along the [111] direction. A transition from semimetal to semiconductor can be realized when Bi is exfoliated into thin film. Figure 1(b) presents the calculated band structure of bulk Bi. The band structure beneath the Fermi level can be described by two filled *s* bands and three filled *p* bands, separated by a band gap of several eV. The *p* bands pass through the Fermi level near T and L points and form hole-type and electron-type pockets which seem very shallow. The electronic DOS shows a steep decrease around the Fermi level (figure 1(c)), which is consistent with the nature of Bi semimetal. As a semimetal, Bi has a very low carrier concentration of $3 \times 10^{17}$ $cm^{-3}$, along with a small effective mass $m^* \sim 0.003\ m_e$ (where $m_e$ is the mass of free electron) [29] and an ultrahigh carrier mobility ($\mu_e \sim 1 \times 10^6$ $cm^2/V$ s at 5 K) [10]. As a result of its large Fermi wavelength and mean free path, Bi has been selected for studies of quantum transport effect and finite-size effect. Moreover, Bi is an important element in the study of topological insulators, such as $Bi_2Se_3$ and $Bi_2Te_3$ [30]. The XMR of Bi was discovered as early as 1999, when Yang et al. observed a large MR of 3.8 × $10^5$% at 5 K and 5 T in single crystal thin films of Bi [5]. The value persists to rise until 40 T and accumulates a huge MR. However, upon further increasing magnetic field, the MR shows a tendency to saturation (figure 1(d)) [31].

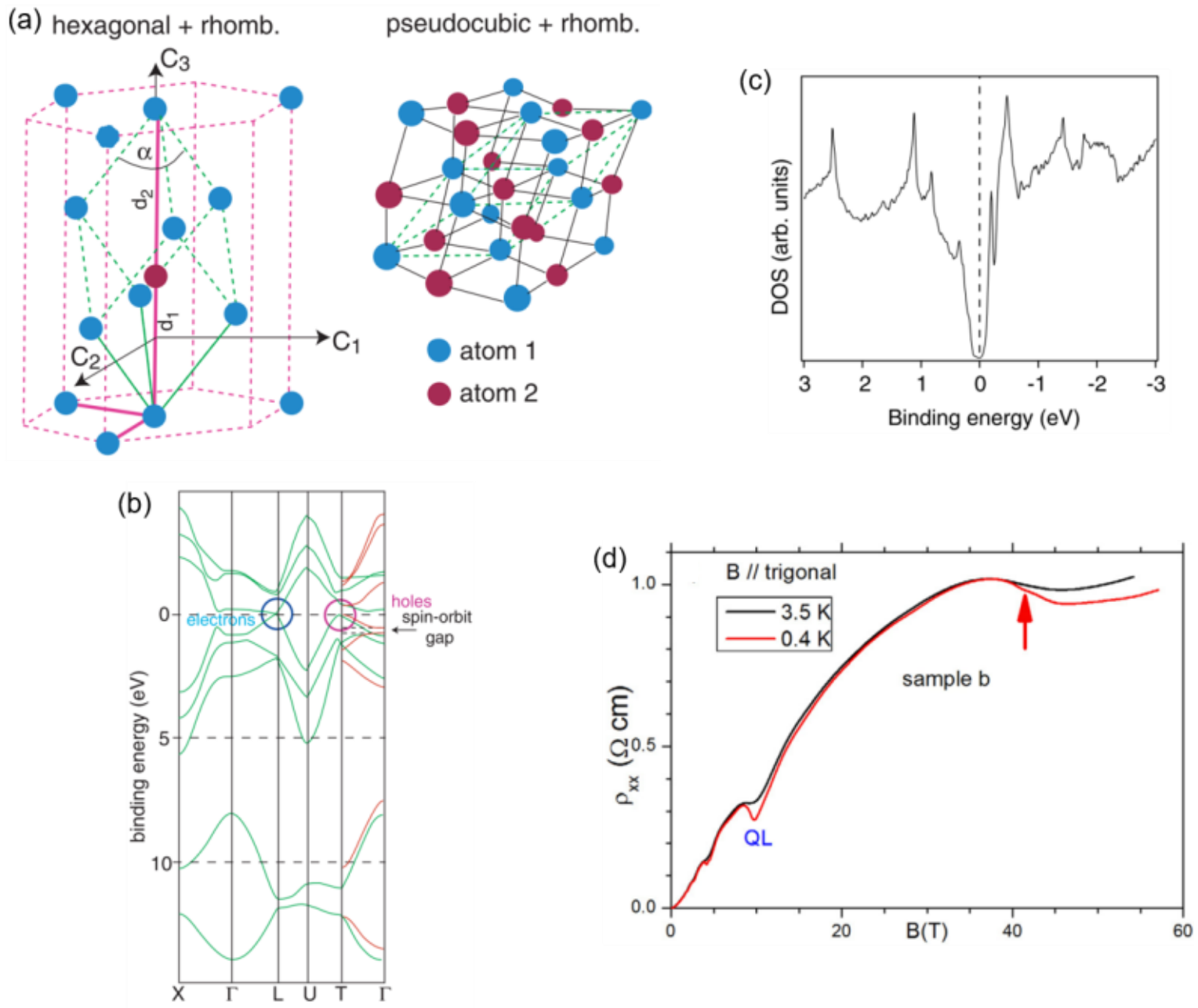

**Figure 1.** (a) Crystal structure of bulk Bi. Left: a rhombohedral unit cell (dashed green lines) superimposed on a hexagonal unit cell (dashed pink lines). Not all the atoms are shown. Blue and red balls represent two atoms in the rhombohedral unit cell. Right: illustration revealing the structure's pseudo cubic character. (b) Calculated energy bands of bulk Bi based on tight binding approximation (green lines) and first-principles calculation method (red lines). (c) DOS of the body near the Fermi level of Bi calculated by tight binding parameters. Adapted from Ref. [29]. (d) Resistivity as a function of magnetic field for Bi single crystal taken at 0.4 K and 3.5 K. Adapted from Ref. [31].

Like Bi, α-As has a rhombohedral crystal structure (*R*-3*m* space group), as shown in figure 2(a). It is also a layered material that is easy to intercalate or exfoliate due to the weak interlayer force. Zhao et al. observed an XMR for α-As single crystal which is as high as 1.5 × $10^7$% at 1.8 K and 9 T (figure 2(b)) and attributed to the electron-hole compensation. A nontrivial Berry phase was revealed by Shubnikov-de Haas (SdH) oscillation measurements and analyses. They also claimed finding of a negative LMR that is usually connected with potential topological character (figure 2(c)) [6].

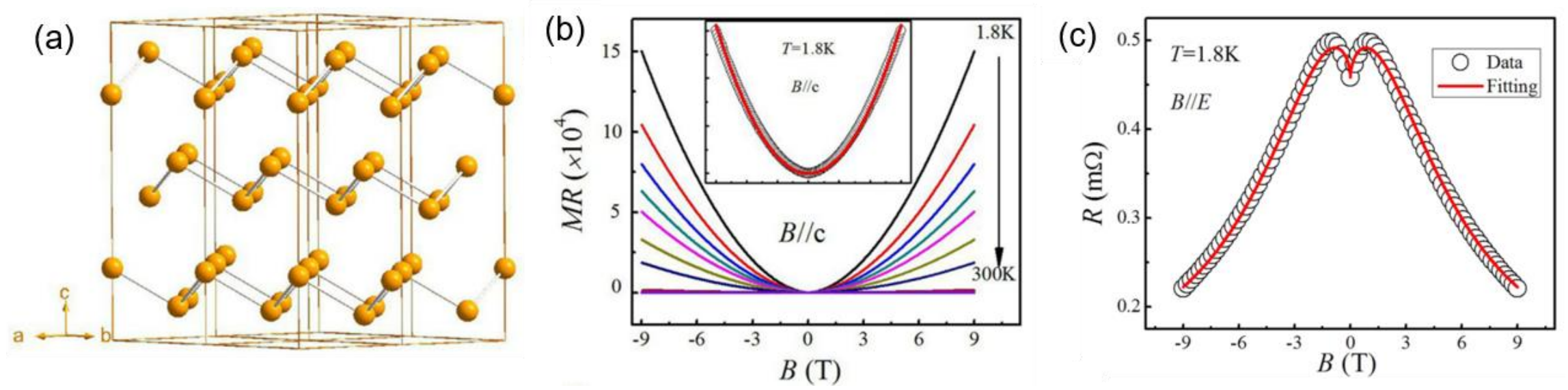

**Figure 2.** (a) Illustrated crystal structure of α-As. (b) Magnetic field dependence of MR taken at various temperatures. Inset: solid line represents a quadratic fit to the MR data at 1.8 K. (c) Resistance measured at 1.8 K with *B* // *I* (open circles) and fitting curve (solid line). Adapted from Ref. [6].

α-Ga belongs to orthorhombic system (*Cmca* space group), as shown in figure 3(a) [32]. Figures 3(b) and 3(c) presents the Fermi surface contour and DOS calculated with spin-orbit coupling (SOC), respectively. The valence bands are derived from a hybrid combination of *s* and *p* states of α-Ga. The DOS shows a minimum at the Fermi level, which indicates α-Ga is a semimetal. The band structures calculated without and with SOC are shown in figures 3(d) and 3(e), respectively. The band crossings appearing at some points suggest α-Ga may possess topological characteristics. Besides, a (superconducting) transition with critical temperature $T_C$ = 0.9 K is demonstrated by specific heat measurements that are dependent on applied magnetic field (figure 3(f)). Since the resistance of bulk α-Ga is very low, which makes it difficult to directly measure the superconducting transition of resistivity, several bars of α-Ga single crystals were connected in series to increase resistance. A sudden drop in resistance was observed at 0.9 K (figure 3(g)), which is consistent with the specific heat measurement. Furthermore, an XMR as large as $1.66 \times 10^6$% was observed at 2 K and 9 T for α-Ga single crystal (figure 3(h)) and the electron-hole compensation was believed to account for the XMR. A small effective mass (0.02-0.05 $m_e$) is revealed by the measurement and analysis of de Haas-van Alphen (dHvA) oscillations [7]. As a pure metal that is in the liquid phase at room temperature and ambient pressure, α-Ga not only has superconductivity and topological properties but also has a prominent XMR, which provides a platform to study superconductivity in topological materials as well as the effect of XMR on these important properties.

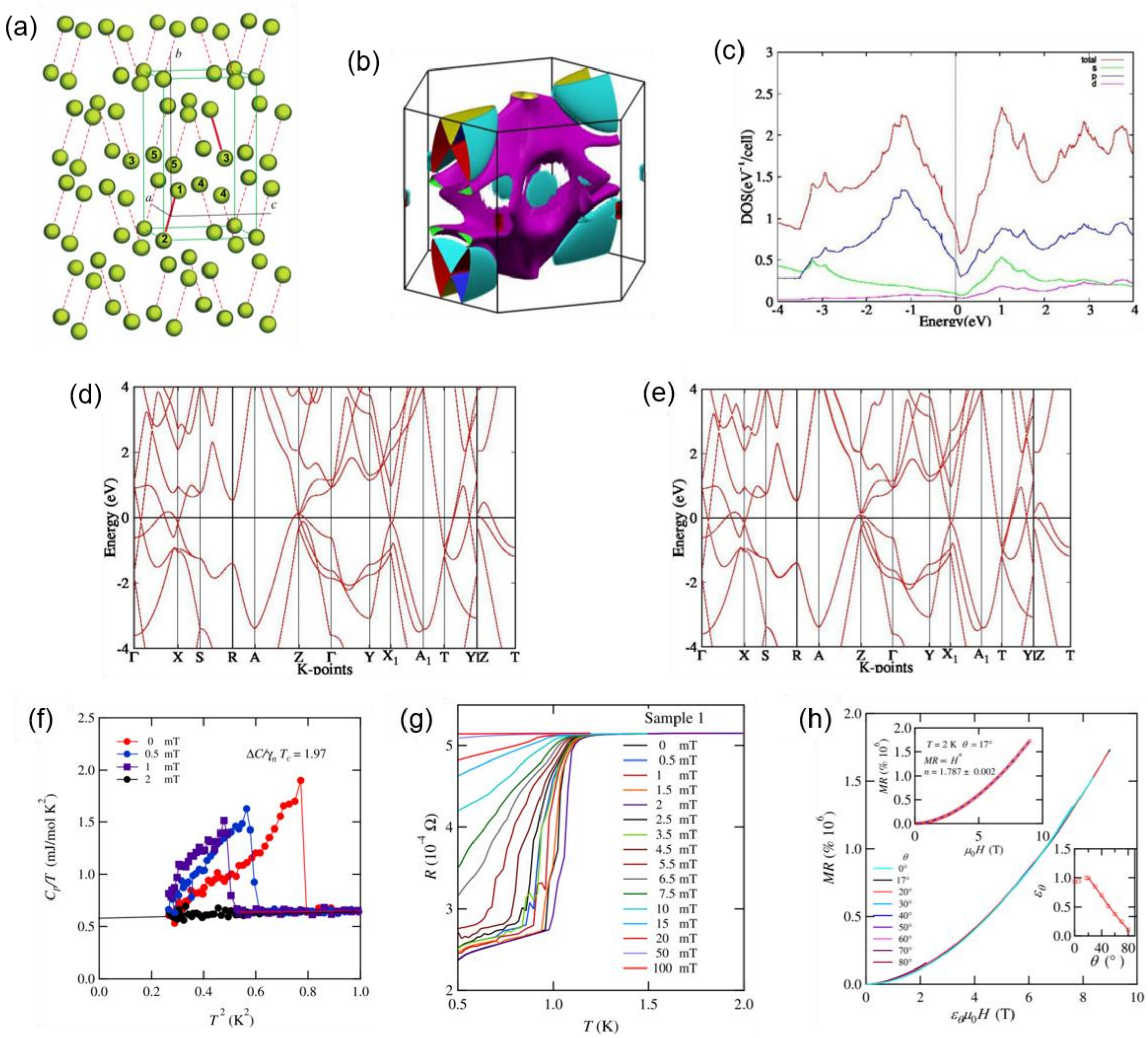

**Figure 3.** (a) Illustrated crystal structure of α-Ga. Adapted from Ref. [32]. (b) Calculated Fermi surface for α-Ga in the first Brillouin zone. (c) Calculated DOS with SOC. (d) Band structures along high symmetry directions calculated without SOC. (e) Band structures calculated with SOC. (f) Specific heat divided by temperature as a function of $T^2$ taken under various magnetic fields. The $\lambda$-like peaks suggest a superconducting transition. (g) Resistance as a function of temperature at various magnetic fields. (h) Scaling behavior of field-dependent MR along different magnetic field directions. Upper inset: MR fitted to a power law MR $\propto B^{1.787}$ at 2 K. Lower inset: scaling factor $\varepsilon_\theta$ as a function of $\theta$. Adapted from Ref. [7].

### *2.2 Binary materials*

Binary materials featuring an XMR usually need to be semimetals or metals in the transport nature. Generally, cations are transition metal ions and anions are group V or VI ions.

According to crystal structure and chemical composition, several categories of binary XMR materials are of special research interest, i.e., XP(As), XSb(Bi), TMDs, etc. Table 2 lists the XMR and measurement conditions of binary materials.

**Table 2.** XMR and measurement conditions of binary materials.

| Material | Measurement conditions | MR (%) | Reference |
|---|---|---|---|
| CrP | 16 mK, 14 T | $2.5 \times 10^3$ | [33] |
| NbP | 1.5 K, 62 T | $8.1 \times 10^6$ | [34] |
| TaP | 2 K, 8 T | $3.28 \times 10^5$ | [11] |
| TaAs | 1.8 K, 9 T | $8 \times 10^4$ | [12] |
| NbAs | 2 K, 9 T | $2.3 \times 10^5$ | [35] |
| NbAs | 1.9 K, 18 T | $4.62 \times 10^5$ | [14] |
| DySb | 1.8 K, 38.7 T | $3.7 \times 10^4$ | [36] |
| GdSb | 2 K, 9 T | $1.21 \times 10^4$ | [37] |
| CeSb | 2 K, 9 T | $1.6 \times 10^6$ | [38] |
| HoSb | 2.65 K, 14 T | $2.42 \times 10^4$ | [39] |
| LaSb | 2 K, 9 T | $9 \times 10^5$ | [40] |
| LaSb | 2 K, 9 T | $1 \times 10^5$ | [41] |
| LaSb | 3 K, 9 T | $4.45 \times 10^4$ | [42] |
| NdSb | 2 K, 8 T<br>1.4 K, 38 T | $3 \times 10^4$<br>$1 \times 10^6$ | [17] |
| NdSb | 2 K, 9 T | $1.2 \times 10^4$ | [43] |
| NdSb | 1.3 K, 60 T | $1.77 \times 10^6$ | [44] |
| ScSb | 2 K, 14 T | $2.8 \times 10^4$ | [45] |
| TmSb | 2.3 K, 14 T | $3.31 \times 10^4$ | [46] |
| YSb | 5 K, 9 T | $1.3 \times 10^5$ | [47] |
| YSb | 2.5 K, 14 T | $3.47 \times 10^4$ | [48] |
| LaBi | 2 K, 14 T | $3.8 \times 10^4$ | [49] |
| LaBi | 2 K, 9 T | $1.5 \times 10^5$ | [50] |
| YBi | 2 K, 9 T | $8 \times 10^4$ | [51] |
| ErBi | 2 K, 9 T | $1.2 \times 10^4$ | [52] |
| InBi | 1.8 K, 7 T | $1.2 \times 10^4$ | [53] |
| $HfTe_2$ | 2 K, 9 T | $1.35 \times 10^3$ | [54] |
| $T_d$-$MoTe_2$ | 2 K, 9 T | 520 | [55] |
| 1T'-$MoTe_2$ | 1.8 K, 14 T | $1.6 \times 10^4$ | [18] |
| $PtTe_2$ | 1.3 K, 62 T | $5 \times 10^4$ | [20] |
| $PtTe_2$ | 2.2 K, 9 T | 800 | [56] |
| $PtTe_2$ | 1.8 K, 9 T | $3.06 \times 10^3$ | [57] |
| $PdTe_2$ | 0.36 K, 30 T | 900 | [19] |
| $WTe_2$ | 1.7 K, 8.27 T | $3.76 \times 10^5$ | [58] |
| $WTe_2$ | 1.6 K, 14 T | $1.06 \times 10^4$ | [59] |
| $WTe_2$ | 0.53 K, 60 T | $1.3 \times 10^7$ | [2] |
| $WTe_2$ | 2 K, 9 T | $1.6 \times 10^5$ | [60] |
| $NiTe_2$ | 2 K, 9 T | $1.25 \times 10^3$ | [61] |
| $MoP_2$ | 2 K, 9 T | $6.5 \times 10^5$ | [21] |
| $WP_2$ | 2.5 K, 63 T | $2 \times 10^8$ | [21] |
| $ZrP_2$ | 2 K, 9 T | $4 \times 10^4$ | [62] |
| α-$WP_2$ | 2 K, 9 T | $4.82 \times 10^5$ | [63] |
| α-$WP_2$ | 2 K, 9 T | $8.74 \times 10^5$ | [64] |
| $MoSi_2$ | 2 K, 14 T | $1 \times 10^7$ | [65] |
| $WSi_2$ | 2 K, 14 T | $1 \times 10^5$ | [66] |
| $HfTe_{4.98}$<br>$HfTe_{4.92}$<br>$HfTe_{4.87}$ | 2 K, 9 T | $1.52 \times 10^3$<br>$2.63 \times 10^4$<br>$6.91 \times 10^3$ | [67] |
| $HfTe_5$ | 2 K, 15 T | $6 \times 10^3$ | [68] |
| $HfTe_5$ | 2 K, 9 T | $4 \times 10^3$ | [69] |
| $HfTe_5$ | 2 K, 9 T | $9 \times 10^3$ | [70] |
| $ZrTe_5$ | 0.3 K, 9 T | $3.3 \times 10^3$ | [71] |
| $ZrTe_5$ | 0.3 K, 13.5 T | 96 | [72] |
| $ZrB_2$ | 2 K, 14 T | $1 \times 10^6$ | [17] |
| $MoAs_2$ | 2 K, 9 T | $2.6 \times 10^3$ | [73] |
| $MoAs_2$ | 1.8 K, 9 T | $3.2 \times 10^4$ | [28] |
| $NbAs_2$ | 2 K, 15 T | $3.032 \times 10^5$ | [74] |
| $TaAs_2$ | 2.5 K, 14 T | $7.3 \times 10^5$ | [75] |
| $TaAs_2$ | 0.5 K, 65 T | $4.0 \times 10^6$ | [76] |
| $NbSb_2$ | 0.4 K, 32 T<br>2 K, 9 T | $4.3 \times 10^6$<br>$1.3 \times 10^5$ | [77] |
| $TaSb_2$ | 2 K, 9 T | $1.5 \times 10^4$ | [78] |
| $TbTe_3$ | 1.8 K, 9 T | $5.6 \times 10^3$ | [79] |
| $PtBi_2$ | 1.8 K, 33 T | $1.12 \times 10^7$ | [80] |
| $PtBi_{2-x}$ | 2 K, 14 T | $2.2 \times 10^3$ | [81] |
| $Cd_3As_2$ | 2 K, 14 T | $3.1 \times 10^3$ | [82] |
| $Cd_3As_2$ | 280 K, 14.5 T | 200 | [83] |
| $Cd_3As_2$ | 4 K, 65 T | $2 \times 10^3$ | [84] |
| $Cd_3As_2$ | 200 K, 14 T | $2 \times 10^3$ | [78] |
| $Cd_3As_2$ | 5 K, 9 T | $1.336 \times 10^3$ | [85] |
| $W_2As_3$ | 1.8 K, 9 T | $3.2 \times 10^4$ | [86] |
| $Bi_2Te_3$ | 340 K, 13 T | 600 | [87] |
| $Na_3Bi$ | 4.6 K, 9 T | 460 | [88] |
| $PdSn_4$ | 1.8 K, 14 T | $7.5 \times 10^4$ | [89] |
| $PtSn_4$ | 1.8 K, 14 T | ac：$5 \times 10^5$<br>b：$1.4 \times 10^5$ | [90] |
| $SiP_2$ | 1.8 K, 31.2 T | $5.88 \times 10^4$ | [91] |
| $TaSe_3$ | 1.9 K, 14 T | $7 \times 10^3$ | [92] |
| $SrAs_3$ | 15 K, 14 T | $8.1 \times 10^3$ | [93] |
| SrPd | 10 K, 4 T | $1 \times 10^5$ | [94] |
| WC | 2 K, 9 T | $7 \times 10^3$ | [95] |

### 2.2.1 XP and XAs

XP and XAs represent four important and intensively-studied semimetals where X stands for Ta or Nb atom. TaP is a body-centered tetragonal material with space group $I4_1md$ (109), as illustrated in figure 4(a) [96]. The structure is non-centrosymmetric, hosting broken space inversion symmetry. Moreover, since there is no magnetic element in TaP, the time reversal symmetry still exists. Figure 4(b) shows the calculated band structures along high symmetry directions. It can be seen that the conduction band and the valence band cross each other near the Σ, Σ' and N points, forming a semimetal ground state [96]. Since the theoretical prediction of a promising Weyl semimetal by Weng et al. in 2014 [97], extensive researches have been carried out. TaP single crystal was successfully prepared and an XMR up to $3.28 \times 10^5$% was observed at 2 K and 8 T (figure 4(c)). By analyzing the Hall resistivity, it was suggested that the material is an electron-hole compensation semimetal [11]. TaAs, an isostructural analogue to TaP, is the first experimentally confirmed Weyl semimetal with broken inversion symmetry. TaAs is also an XMR material, with an XMR as high as $8 \times 10^4$% appearing at 1.8 K and 9 T as well as an ultrahigh carrier mobility, namely $\mu_e \approx 1.8\times10^5$ cm$^2$/V s at 10 K. Through transport measurements, it was found that there are both n-type and p-type carriers in TaAs and they almost compensate each other [12].

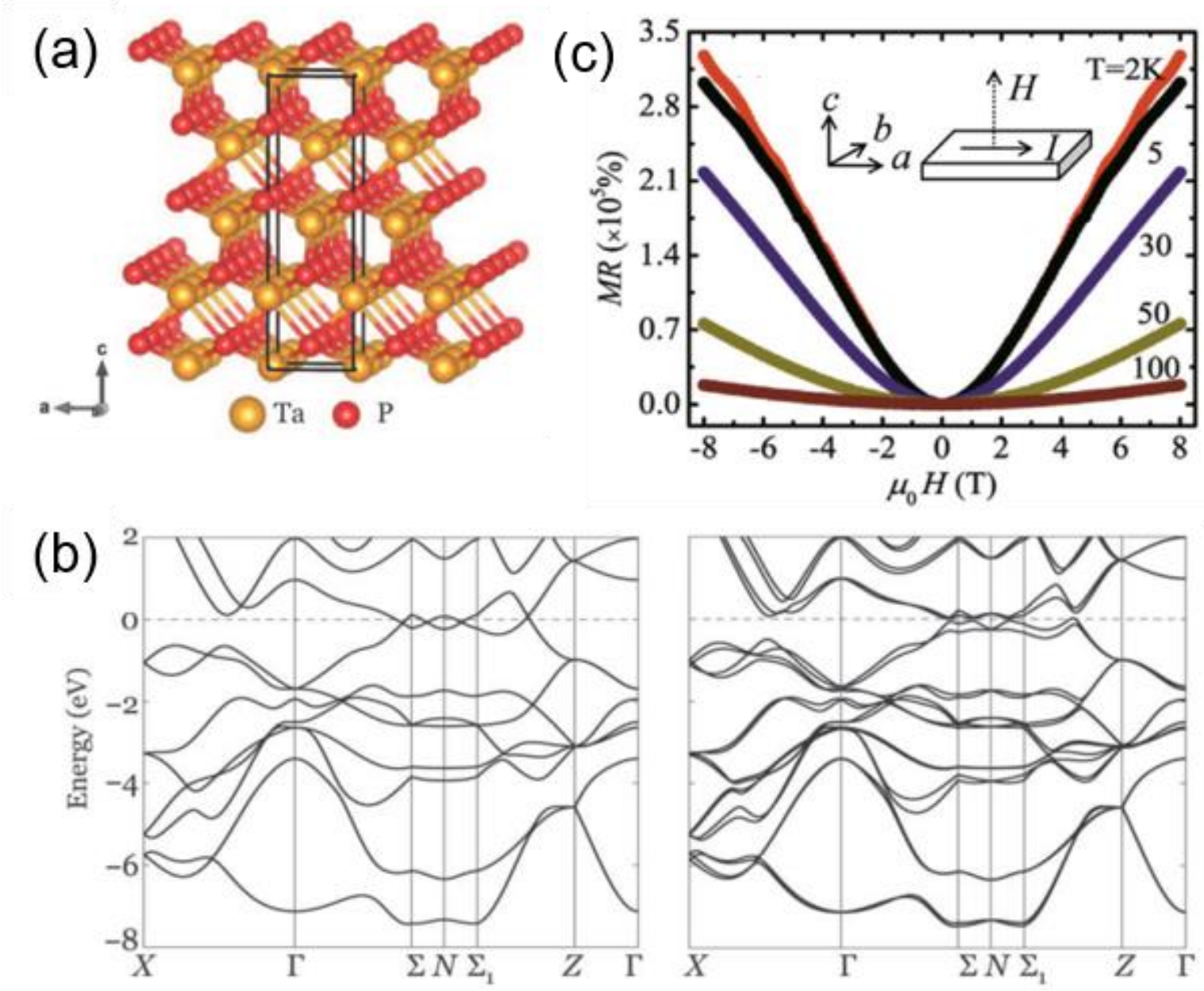

**Figure 4.** (a) Illustrated crystal structure of TaP. (b) First-principles band structures calculated for bulk TaP without and with SOC, respectively. Adapted from Ref. [96]. (c) Magnetic field dependence of MR below 100 K. Adapted from Ref. [11].

### *2.2.2 XSb and XBi*

XSb and XBi are a big family, in which X covers nearly all the rare earth elements (Y, Sc and lanthanides). XMR has been widely observed in XSb and XBi compounds [36-40, 44-47, 50-53, 98]. These materials are important candidates for exploring the origin of XMR due to their simple rock salt cubic structure. The research interest in nonmagnetic monopnictides focused on the so-called "resistance plateau" arising at low temperatures, while the interest in magnetic monopnictides was to search for magnetic topological semimetals. Here we take LaSb, LaBi and NdSb as an example for the nonmagnetic and magnetic monopnictides, respectively.

LaSb has a face-centered cubic structure with space group *Fm-3m* (figure 5(a)) [15]. Figures 5(b) and 5(c) show the calculated Fermi surfaces and band structures for LaSb, respectively. There are two hole-like pockets crossing the Fermi surface near Γ point and an electron-like pocket crossing the Fermi surface near X point [41]. The electronic structure of LaSb is the same along $k_x$, $k_y$ and $k_z$ directions in Brillouin zone, consistent with its simple cubic structure. Zeng et al. calculated the ratio of electron-hole carrier density from quantum oscillation data, i.e., $n_e / n_h = 0.998$, which indicates that LaSb is a compensated semimetal [15]. In 2015, F. F. Tafti et al. reported a resistance plateau for LaSb when breaking the time inversion symmetry (figure 5(e)), which was claimed as a common characteristic of topological insulator. The logic is like this. The resistance increases with the decrease in temperature. At low temperature, the resistance tends to saturate and a plateau appears. It is believed that this is due to the metal-insulator phase transition. For a topological insulator, the surface state is metallic and the body state is insulating, which leads to the increase in resistance with the decreasing temperature. This characteristic depends on the protection of time reversal symmetry. However, this claim has been challenged by other researches [15, 99, 100]. It is now believed that, different from TaP and TaAs, LaSb is topologically trivial, although it has a linear dispersive band [15]. Transport measurement reveals a high carrier mobility, $\mu_e = 1.1 \times 10^4$ cm$^2$/V s [42], and an XMR up to $9 \times 10^5$% at 2 K and 9 T (figure 5(d)), which is attributed to the electron-hole compensation [40].

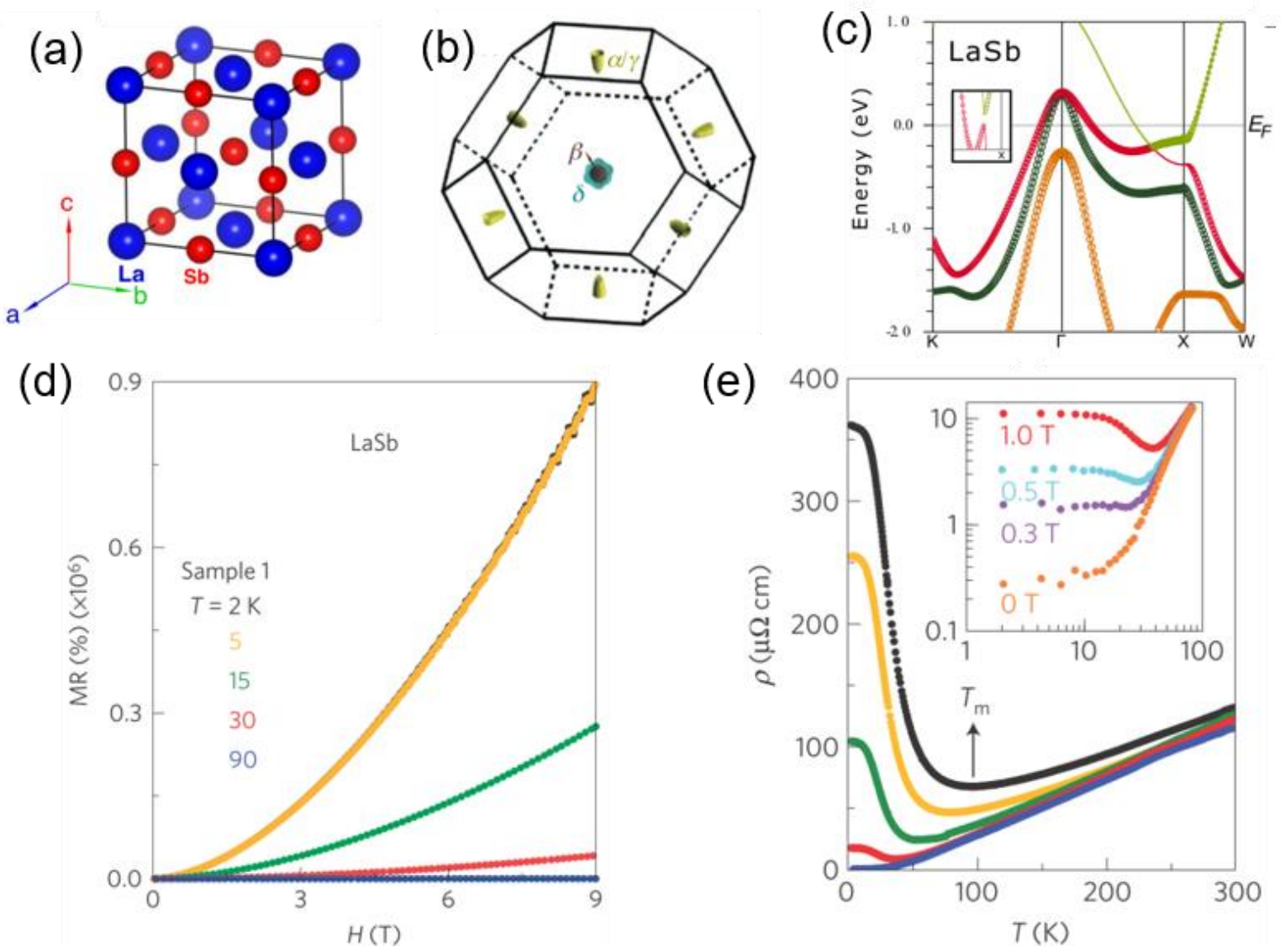

**Figure 5.** (a) Schematic illustration of crystal structure for LaSb. (b) Calculated 3D Fermi surfaces with the MBJ potential. Adapted from Ref. [15]. (c) Band structures of LaSb along high symmetry directions. Adapted from Ref. [41]. (d) MR as a function of magnetic field taken at different temperatures for LaSb. (e) Resistivity as a function of temperature under various magnetic fields, showing a plateau-like structure at low temperatures. Adapted from Ref. [40].

An XMR up to 3.8 × $10^4$% at 2 K and 14 T was also observed in LaBi, as well as a field-induced resistivity upturn and plateau like LaSb [49]. Through transport measurements, band structure calculation and SdH quantum oscillation analysis, Sun et al. connected the XMR with the electron-hole compensation and ultrahigh carrier mobility ($\mu_e$ = 2.6 × $10^4$ $cm^2$/V s and $\mu_h$ = 3.3 × $10^4$ $cm^2$/V s at 2 K) [49]. Kumar et al. further discovered the anisotropic behavior of XMR in LaBi. At 2 K and 9 T, the XMR is 1.5 × $10^5$% when the magnetic field is along the [101] direction, whereas the XMR is 8 × $10^4$% along the [001] direction [50].

NdSb is an antiferromagnetic (AFM) semimetal (i.e., NiO type) with the same rock salt cubic lattice [44]. Figures 6(a) and 6(b) show the calculated Fermi surfaces and band structures of NdSb in the AFM state. The Fermi surface is characteristic of Fermi pockets, that is, the hole pocket in the zone center and the electronic pocket in the zone corner, and the band inversion occurs near X point along U1-X line [17]. The theoretical calculations are consistent with the Fermi surface mapping and dispersion mapping data revealed by ARPES (figures 6(c) and 6(d)) [16], and jointly suggest a topological semimetal state in NdSb. Topological semimetals previously confirmed by theory and experiment are most nonmagnetic materials. However, there are few studies on the magnetic materials that lack the time reversal symmetry. The main reason is that the measurement will be disturbed due to the existence of magnetic domains. Wang et al. observed a negative LMR phenomenon in NdSb that can be well fitted to a chiral anomaly model (figures 6(e) and 6(f)), and confirmed that NdSb is a topological semimetal which lacks the time reversal symmetry. Meanwhile, an XMR as high as 1.6 × $10^6$% was observed at 0.39 K and 38 T (figure 6(g)) [17]. At 1.3 K and 60 T, the XMR was recorded as 1.77 × $10^6$% (figure 6(h)) [44]. By analyzing the SdH quantum oscillation data, the XMR in NdSb was ascribed to the perfect electron-hole compensation and specific orbital structure.

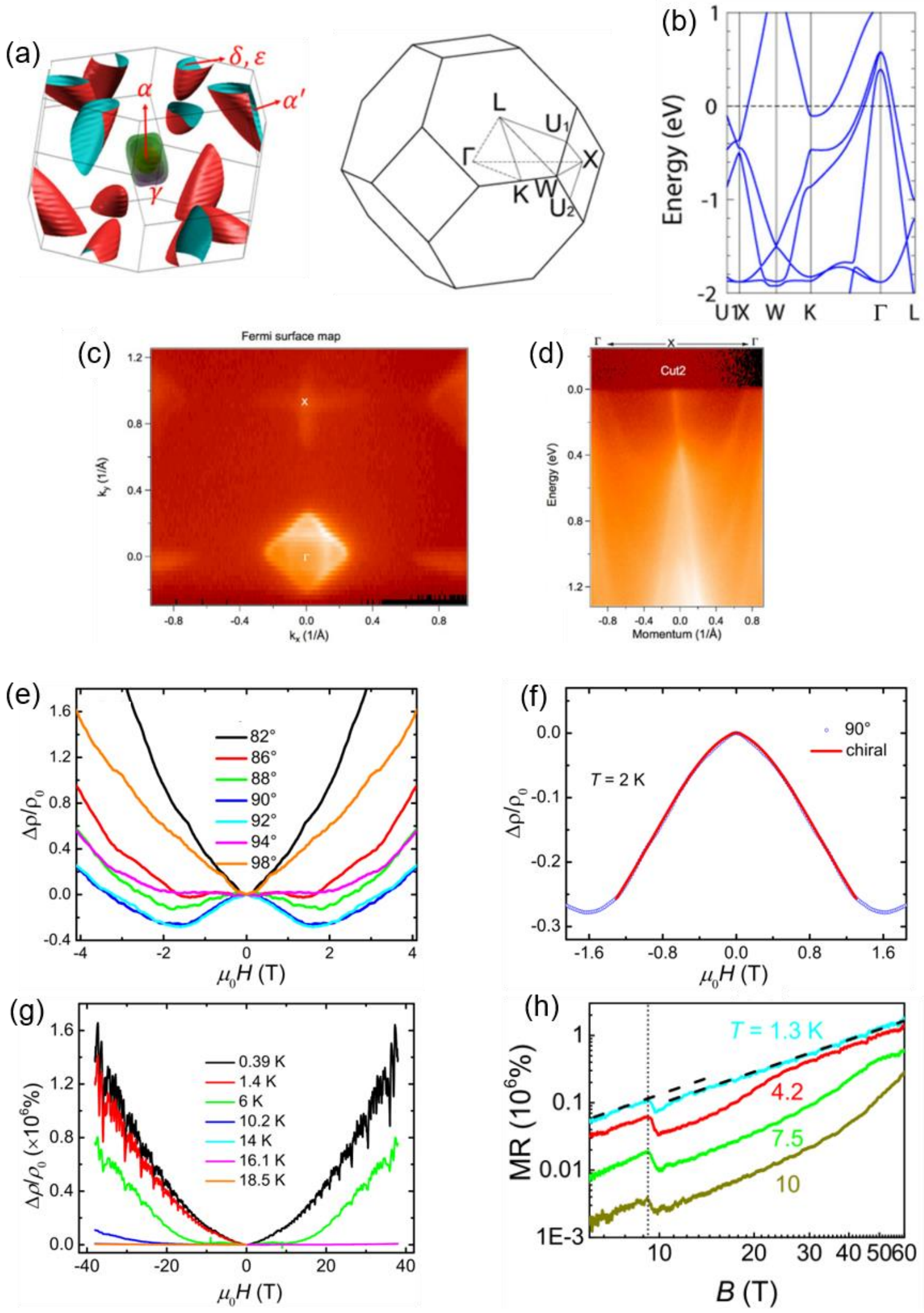

**Figure 6.** (a) Left: Fermi surfaces calculated for AFM state (spin up and spin down are the same) of NdSb. Right: high-symmetry points in the Brillouin zone. U1 and U2 are adjacent equivalent points for crystalline symmetry. (b) Band structures plotted along different *k* paths for AFM state. (c) ARPES measured Fermi surface mapping along the X-Γ-X direction with incident photon energy 50 eV at a temperature of 20 K. (d) Dispersion map along Γ-X-Γ. (e) MR measured at different angles around $\theta$ = 90° (from 82° to 98°) at 2 K. $\theta$ = 90◦ means that *B* is parallel to *I*. (f) Fitting results of the negative LMR data at 2 K and θ = 90◦. Solid line represents the fitted curve with chiral anomaly formula. (g) High-field MR measured at different temperatures, with *B* applied perpendicular to *I* and sample plane. (h) Logarithmic plot of MR versus applied magnetic field up to 60 T with $B \perp I$ at various temperatures. (a), (b) and (e)-(g) are adapted from Ref. [17]. (c) and (d) are adapted from Ref. [16]. (h) is adapted from Ref. [44].

### *2.2.3 TMDs*

Transition metal dichalcogenides are important layered compounds with a wide range of transport properties, extending from semiconductors to semimetals and metals, also involving topological novelty. XMR commonly appears in tellurides that are semimetals or metals in nature. Typical representatives include $WTe_2$, $MoTe_2$, $PdTe_2$ and $PtTe_2$ [2, 18-20, 101].

$PtTe_2$ crystallizes in the $CdI_2$-type trigonal structure (1T) with space group *P*-3*m*1 (figure 7(a)) [102]. As a layered structure material, the layers are connected by vdW force which is relatively weak. Hence, $PtTe_2$ crystals can be easily cleaved or stripped by physical or chemical force. Moreover, a vdW epitaxial growth of $PtTe_2$ crystal (4.2 nm) with highly crystallized atomic layer can be realized on mica [56]. As shown in figure 7(b), ARPES measurements reveal a pair of strongly tilted Dirac cones in $PtTe_2$, which is consistent with the first-principles calculations. As presented in figure 7(c), at the Γ point the band inversion between $\Gamma_4^+$ and $\Gamma_4^-$ forms a topologically nontrivial gap, giving rise to the surface states of the gapped cone structure. The bulk Dirac cone is formed by two Te-*p* valence bands (red color). Because the two bands belong to different irreducible representations, hybridization between each other is forbidden and the band crossing occurs [102]. The ARPES measurements and theoretical calculations corporately confirm that $PtTe_2$ is a type-II Dirac semimetal [102]. Figure 7(d) shows the XMR of $PtTe_2$ up to 5 × $10^4$% at 1.3 K and 62 T [20].

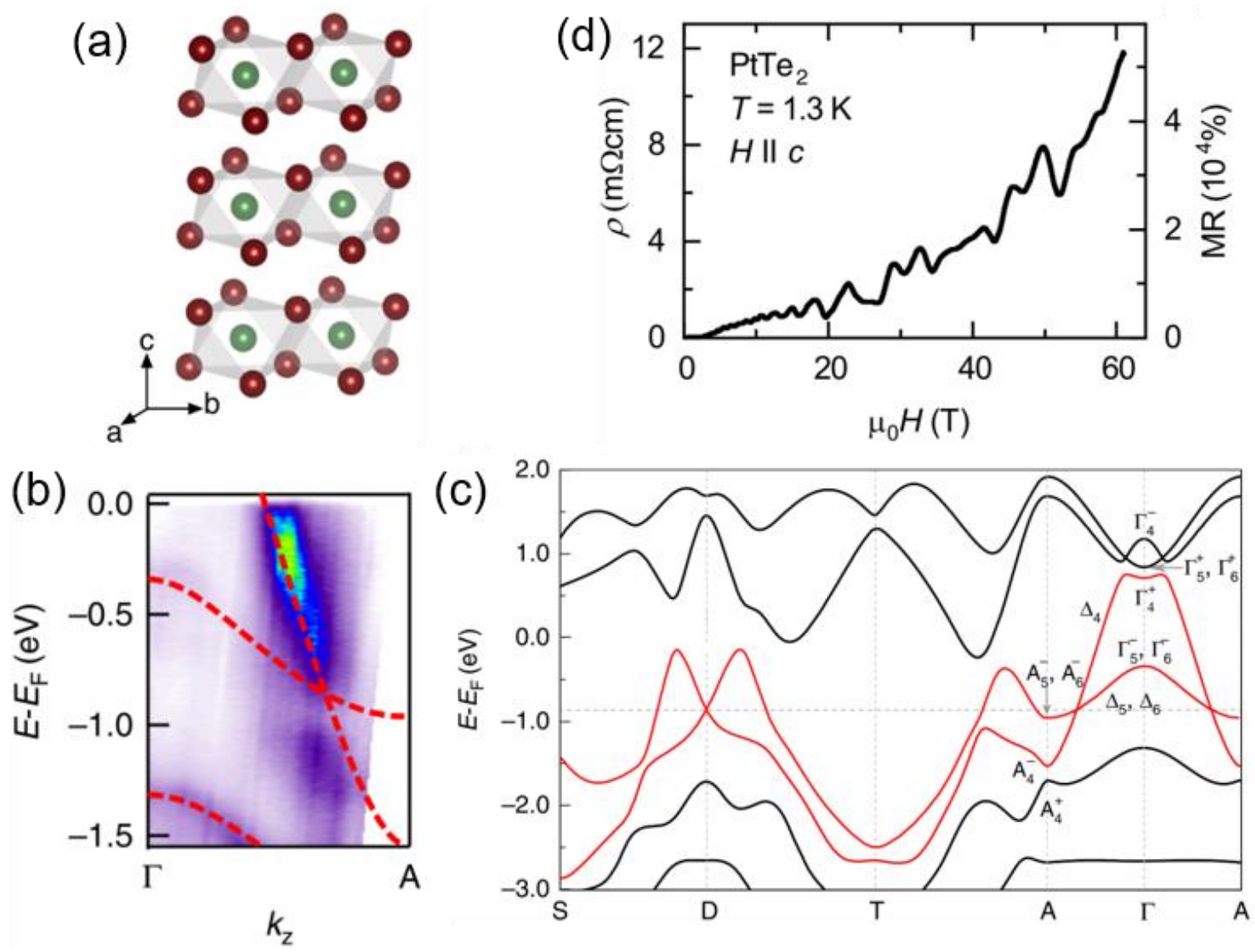

**Figure 7.** (a) Schematic illustration of crystal structure for $PtTe_2$. (b) Measured dispersion at $k_{//} = 0$. Red dashed lines are calculated dispersions for comparison. (c) Calculated band structures. Adapted from Ref. [102]. (d) Resistivity and MR as a function of magnetic field up to 62 T taken at 1.3 K for $PtTe_2$ with magnetic field parallel to the *c* axis. Adapted from Ref. [20].

$PdTe_2$, structurally akin to $PtTe_2$, is another type-II Dirac semimetal [103]. It is more interesting that superconductivity has been observed in pure $PdTe_2$ [19] and Cu-intercalated $PdTe_2$, i.e., $Cu_{0.05}PdTe_2$ [104], which provides an important platform to study the coexistence and correlation of topological surface state and bulk superconductivity. This is also an important pathway to realization of potential topological superconductivity. Wang et al. reported an XMR of 900% appearing at 0.36 K and 30 T [19].

### *2.2.4 $WP_2$ and $WSi_2$*

Other than TMDs, some compounds with the $AB_2$ formula also possess a significant XMR and intriguing topological characteristics. $WP_2$ has an orthorhombic structure with space group $Cmc2_1$ (figure 8(a)). As a three-dimensional (3D) transition metal diphosphide, $WP_2$ is considered to be a type-II Weyl semimetal with stable Weyl points, investigated via transport measurements, ARPES and first-principles calculations. The electronic structure is directly measured by ARPES on the (010) surface of $WP_2$, showing no open Fermi arc, because the projections of a pair of chiral opposite Weyl nodes overlap with each other on the (010) surface. Figure 8(c) shows the Fermi surface mapping, in which $\bar{A}$-$\bar{X}$ and $\bar{Z}$-$\bar{\Gamma}$ directions cross the hole and electron pockets, respectively. In addition, the energy dispersion along high symmetry lines confirms the semimetal nature of $WP_2$ (figure 8(d)). Further studies on other surfaces, especially on the (001) surface, are needed to identify the possible arc states. The XMR in $WP_2$ can be as high as $2 \times 10^8$% at 2.5 K and 63 T (figure 8(e)), which is attributed to a combination of electron-hole compensation ($n_h = 1.5\times10^{20}$ cm$^{-3}$, $n_e = 1.4\times10^{20}$ cm$^{-3}$) , high mobility ($1.65\times10^6$ cm$^2$/V s) and extremely large RRR (24,850) [21].

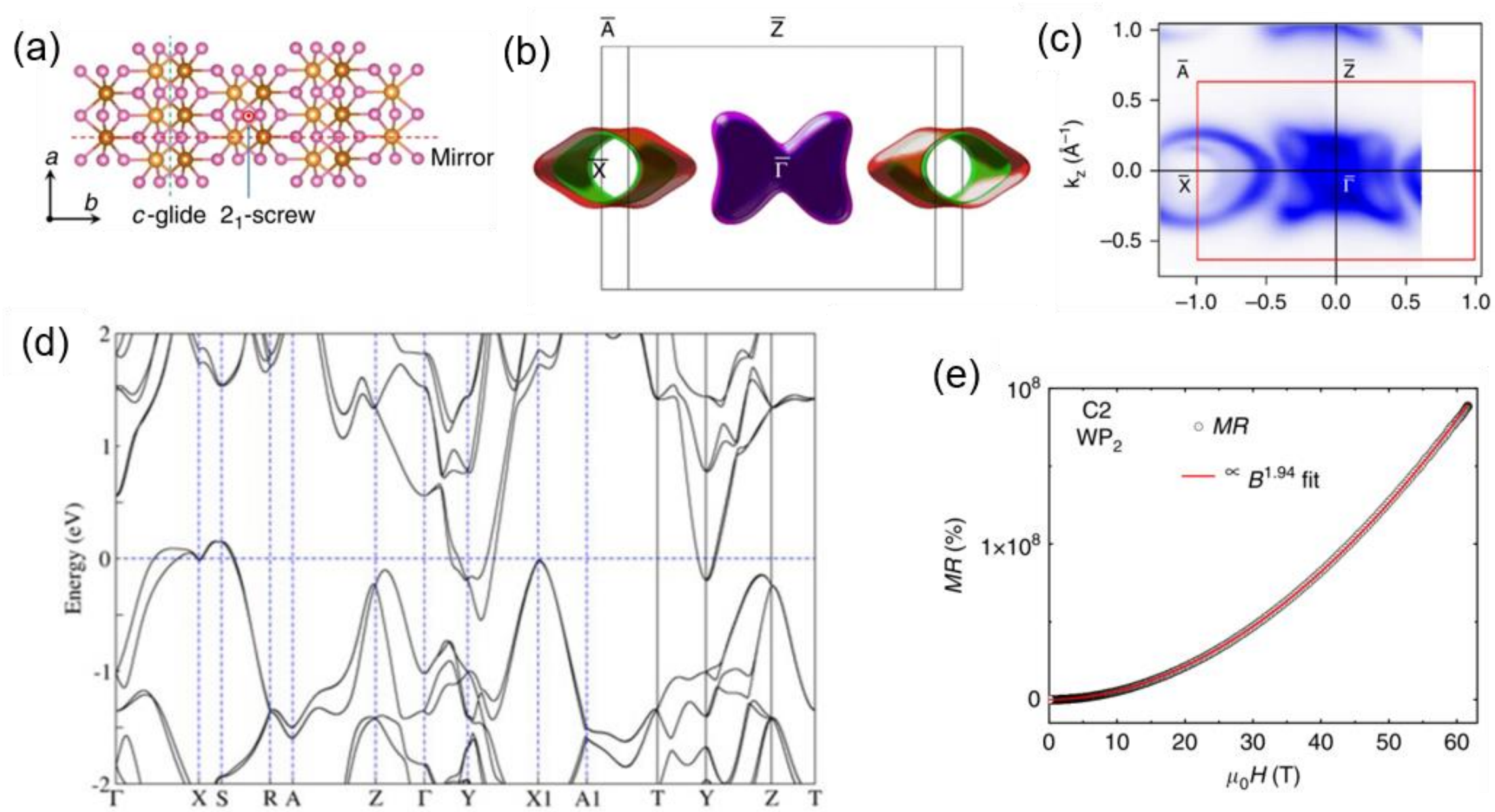

**Figure 8.** (a) Illustrated crystal structure of $WP_2$. (b) Projection of calculated Fermi surface in the *ac*-plane. Spaghetti-like open hole Fermi surfaces located around X point in the Brillouin zone, extending along the *b* axis. Bow-tie-like closed electron Fermi surfaces located around Y point in the Brillouin zone. (c) Fermi surface cross section of $WP_2$ along the *b* axis from ARPES measurements showing good correspondence with the calculated Fermi surface. (d) Energy dispersions along high symmetry directions. (e) MR taken at 2.5 K in a pulsed magnetic field up to 63 T. Red line shows a near-perfect parabolic fit. Adapted from Ref. [21].

The crystal structure of $WSi_2$ is tetragonal with space group *I4/mmm* (figure 9(a)). As shown in figure 9(b), in the band structures the electron and hole pockets are centered at the high symmetry Z point and Γ point, respectively. Through transport measurement, dHvA quantum oscillation analysis and electronic structure calculations, the XMR of $2 \times 10^5$% (2 K, 14 T) is believed to originate from extremely large carrier mobility ($\mu_h = 3.9 \times 10^4$ cm$^2$/V s, $\mu_e = 3.4 \times 10^4$ cm$^2$/V s) and near compensated charge carriers. Furthermore, the dHvA studies reveal a nontrivial Berry phase, suggesting possible topological properties [66]. In the same work, $MoSi_2$ with the same crystal structure is also investigated, and an XMR up to $10^7$% at 2 K and 14 T is reported, which is again connected with carrier compensation ($n_h/n_e = 0.954$) and ultrahigh carrier mobility ($\mu_h = 4.5 \times 10^5$ cm$^2$/V s, $\mu_e = 2.7 \times 10^5$ cm$^2$/V s) [65].

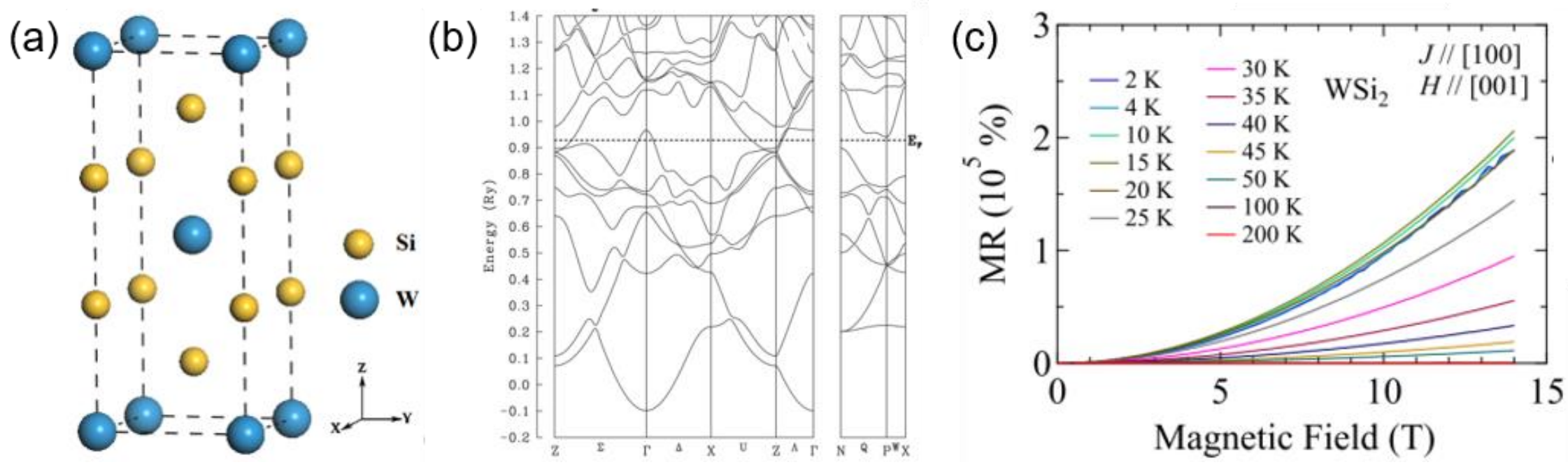

**Figure 9.** (a) Schematic crystal structure of $WSi_2$. Adapted from Ref. [105]. (b) Band structure calculations. (c) Field dependence of transverse MR taken at different temperatures with the field along [001]. Adapted from Ref. [66].

### *2.2.5 $ZrTe_5$ and $HfTe_5$*

$ZrTe_5$ and $HfTe_5$ are another pair of binary compounds that show both semimetal transport and XMR characteristic, though some researchers consider $ZrTe_5$ is a topological insulator [106-108]. As illustrated in figure 10(a), $ZrTe_5$ crystallizes in a layered orthorhombic structure with space group *Cmcm*. $ZrTe_5$ has been famous for its prominent thermoelectric effect, as well as sometimes elusive topological properties. Observation of negative LMR (figure 10(c)) by Zheng et al. is linked to chiral anomaly and nontrivial Berry phase, which suggests their $ZrTe_5$ sample is a possible Dirac semimetal [109]. This is in correspondence with the ARPES measurement shown in figure 10(b). However, since the Dirac point is not occupied, the ARPES is not able to give a complete picture for the electronic structure near the Dirac point, and the existence of a small mass gap cannot be ruled out [110]. In addition, Chen et al. performed the magnetoinfrared spectroscopy for $ZrTe_5$ and observed a Landau level splitting, which was considered to be the transition from Dirac semimetal to nodal-line semimetal [22]. Furthermore, the effect of hydrostatic pressure on the magnetotransport of $ZrTe_5$ was studied. The resistivity was found to decrease with the increasing pressure. At ambient pressure, 0.3 K and 9 T, the MR of $ZrTe_5$ reaches 3300%, which undergoes a rapid drop upon further increasing the field (figure 10(d)). At 2.5 GPa, the MR decreases to 230% [23]. An XMR is also observed in $HfTe_5$ with the same crystal structure. As a well-known layered thermoelectric material [111], $HfTe_5$ has been recently predicted to be a large gap quantum spin Hall insulator [112]. Confusion arises again on $HfTe_5$. $HfTe_5$ is also regarded as a 3D Dirac semimetal [69]. A preliminary conjecture is based on the possible slight difference in sample stoichiometry. A moderate XMR of $HfTe_5$, $9 \times 10^3$% at 2 K and 9 T, is attributed to high carrier mobility ($\mu_{avg} = 2.8\times10^4$ $cm^2/V$ s) [70].

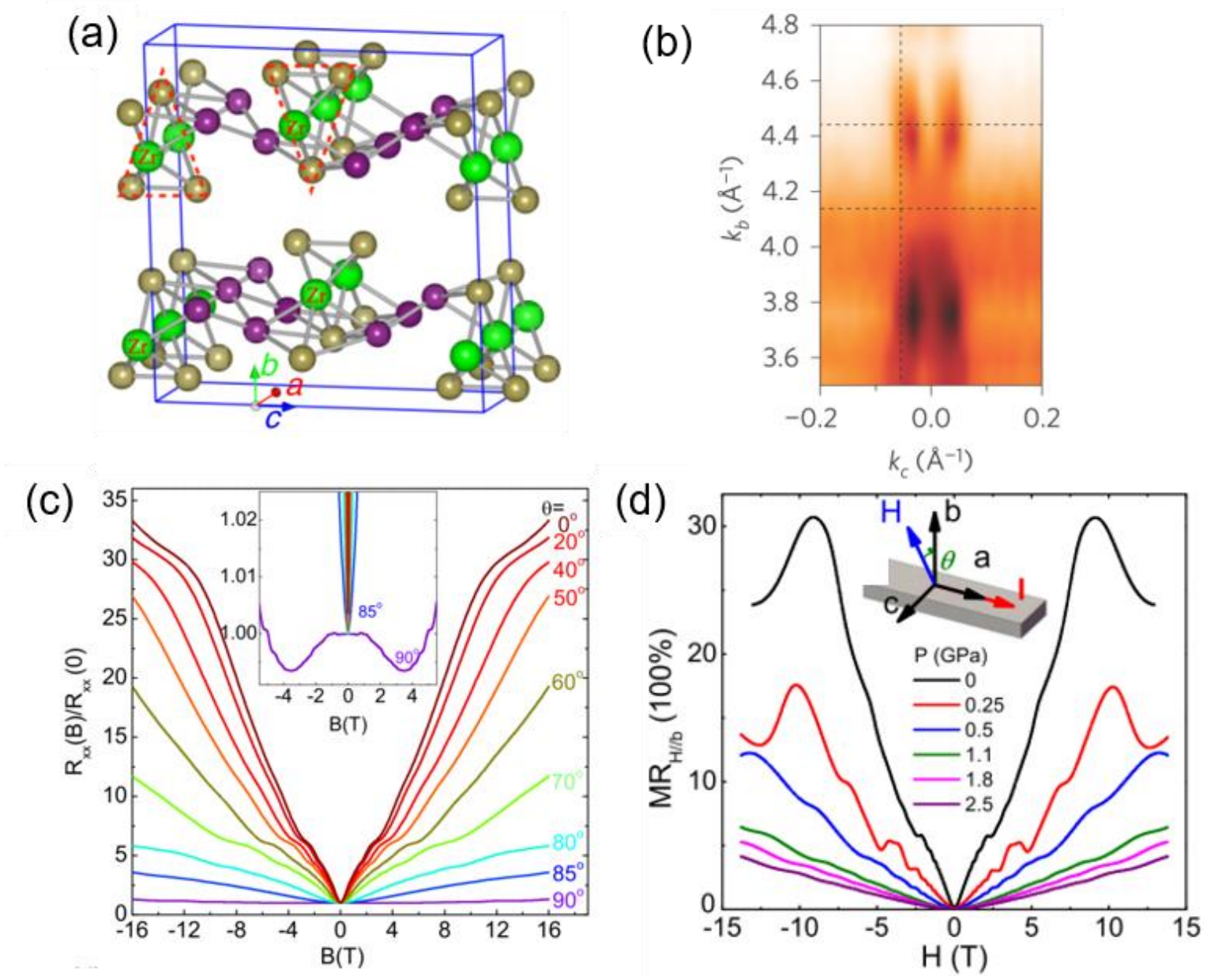

**Figure 10.** (a) Schematic crystal structure of $ZrTe_5$. Adapted from Ref. [112]. (b) Out-of-plane Fermi surface contour as a function of the in-plane momentum perpendicular to the chain, $k_c$, and the momentum perpendicular to the surface, $k_b$. Adapted from Ref. [110]. (c) Angular dependence of LMR by tilting $B$ in the *ab* plane measured at 2 K, where a dc current is applied along the *a* axis. Adapted from Ref. [109]. (d) MR of $ZrTe_5$ taken at 0.3 K and various pressures with $B$ // $b$ and $I$ // $a$. Inset is a schematic of measurement geometry. Adapted from Ref. [23].

### *2.2.6 Other*

In addition, there are still some binary materials that show an XMR but cannot be listed into the above categories. For instance, an XMR up to $10^6$% was observed at 2 K and 14 T for $ZrB_2$, a predicted topological nodal-line semimetal. Combining transport measurement, band structure calculations and SdH quantum oscillation analysis, Wang et al. attributed the XMR in $ZrB_2$ to the electron-hole compensation ($n_e/n_h = 1.003$) and high carrier mobility ($\mu_e = 1.66 \times 10^4$ $cm^2/V$ s, $\mu_h = 1.64 \times 10^4$ $cm^2/V$ s) [113]. In $TaAs_2$, an XMR of $4.0 \times 10^6$% at 0.5 K and 65 T was observed [76]. And for $NbSb_2$, Wang et al. reported an XMR of $4.3 \times 10^6$% at 0.4 K and 32 T, which was related to high mobility [77].

### 2.3 Ternary materials

Ternary materials with a significant XMR that have been intensively studied in recent years involve half-Heusler alloys, $Co_3Sn_2S_2$, ZrSiS, ZrSiSe, etc. Most of them are of particular significance in topological research such as magnetic topological semimetals and topological nodal-line semimetals. Table 3 lists the MR and measurement conditions of some ternary materials.

**Table 3.** MR and measurement conditions of ternary materials.

| Materials | Measurement conditions | MR (%) | Reference |
|---|---|---|---|
| GdPtBi | 2 K, 14 T | 285 | [25] |
| ZrSiSe | 4.2 K, 60 T | $1 \times 10^8$ | [24] |
| ZrSiSe | 2.3 K, 14 T | $3 \times 10^4$ | [114] |
| ZrSiS | 2 K, 9 T | $1.4 \times 10^5$ | [115] |
| $Co_3Sn_2S_2$ | 2 K, 12 T | 208 | [116] |
| $Co_3Sn_2S_2$ | 2 K, 14 T | 139 | [117] |
| $Co_3Sn_2S_2$ | 2 K, 9 T | 220 | [26] |
| $Co_3Sn_2S_2$ | 2 K, 6 T | 215 | [118] |
| $TaPdTe_5$ | 2.1 K, 51.7 T | $9.5 \times 10^3$ | [119] |
| CaCdSn | 4 K, 12 T | $7.44 \times 10^3$ | [120] |
| YbCdGe | 3 K, 12 T | $1.14 \times 10^3$ | [121] |

Half-Heusler alloys have attracted much attention for their thermoelectricity, magnetism and superconductivity. GdPtBi is a representative half-Heusler alloy which has been recently studied as a topological material. GdPtBi crystallizes in a cubic MgAgAs-type structure with space group $F\text{-}43m$ (figure 11(a)). The electronic band structures are shown in figure 11(b). There are two parabolic bands at the Γ point touching each other, and there is a three-fold point along the Γ-L line [122]. Hirschberger et al. reported a negative LMR caused by chiral anomaly (figure 11(c)), and believed that GdPtBi is a Weyl semimetal [123]. Furthermore, Zhang et al. retrieved information of nontrivial Berry phase, negative LMR and giant planar Hall effect in GdPtBi (figures 11(d)-(f)) from their transport measurements, which confirms the topological state of GdPtBi. In addition, an MR of 285% appears at 2 K and 14 T (figure 11(f)), which is just close to the lower limit of XMR [25]. GdPtBi and $Co_3Sn_2S_2$ are not listed as XMR materials (see Table 3).

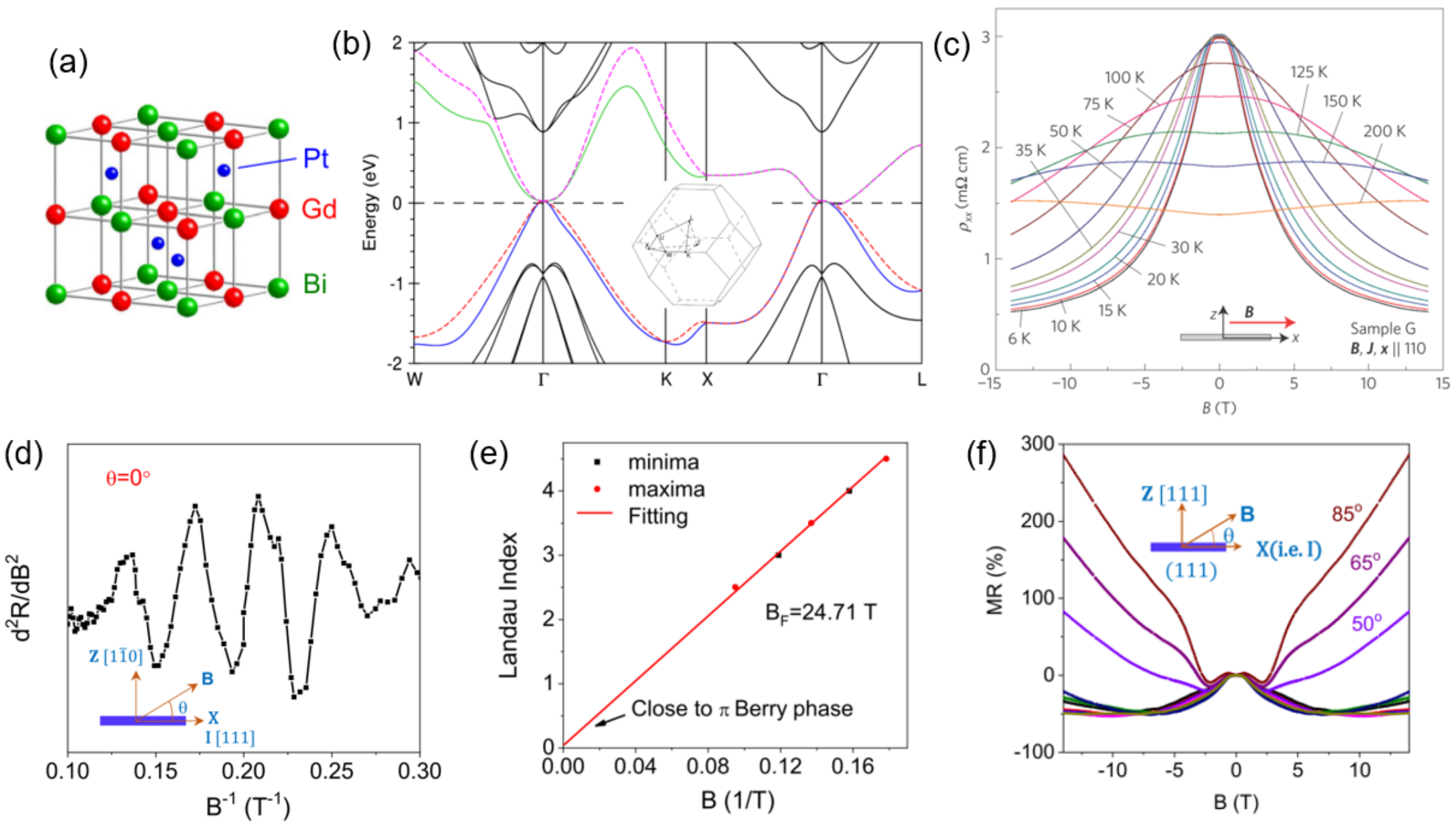

**Figure 11.** (a) Schematic crystal structure of GdPtBi. (b) Band structures of GdPtBi. Inset is the first Brillouin zone showing high symmetry points. Adapted from Ref. [122]. (c) $\rho_{xx}(B)$ shows a negative LMR at 6 K with a bell-shaped profile, which remains resolvable to >150 K. Adapted from Ref. [123]. (d) SdH oscillation and (e) Landau index $n$ as a function of $1/B$ at $\theta = 0$. (f) Variation of MR measured at 2 K and various angles. Adapted from Ref. [25].

ZrSiSe has a tetragonal structure with space group *P*4/*nmm* (figure 12(a)). The calculated electronic band structures are shown in figure 12(b). The conduction band and valence band (orange and violet curves) intersect at the Fermi level. The two bands show Dirac-like dispersion at specific points of Brillouin zone, and the Dirac nodes occur near the Fermi level. As a representative nodal-line semimetal, ZrSiSe not only has the common characteristics of Dirac semimetals, i.e., ultrahigh carrier mobility ($\mu_{avg} = 1.89 \times 10^5$ cm$^2$/V s) and distinct SdH and dHvA quantum oscillations, but also has an outstanding XMR. The supreme XMR up to $10^8$% at 4.2 K and 60 T was attributed to the electron-hole compensation ($n_e/n_h = 0.961$), as shown in figure 12(c) [24]. ZrSiS, a structural analogue to ZrSiSe, exhibits an XMR as high as $1.4 \times 10^5$% at 2 K and 9 T [115].

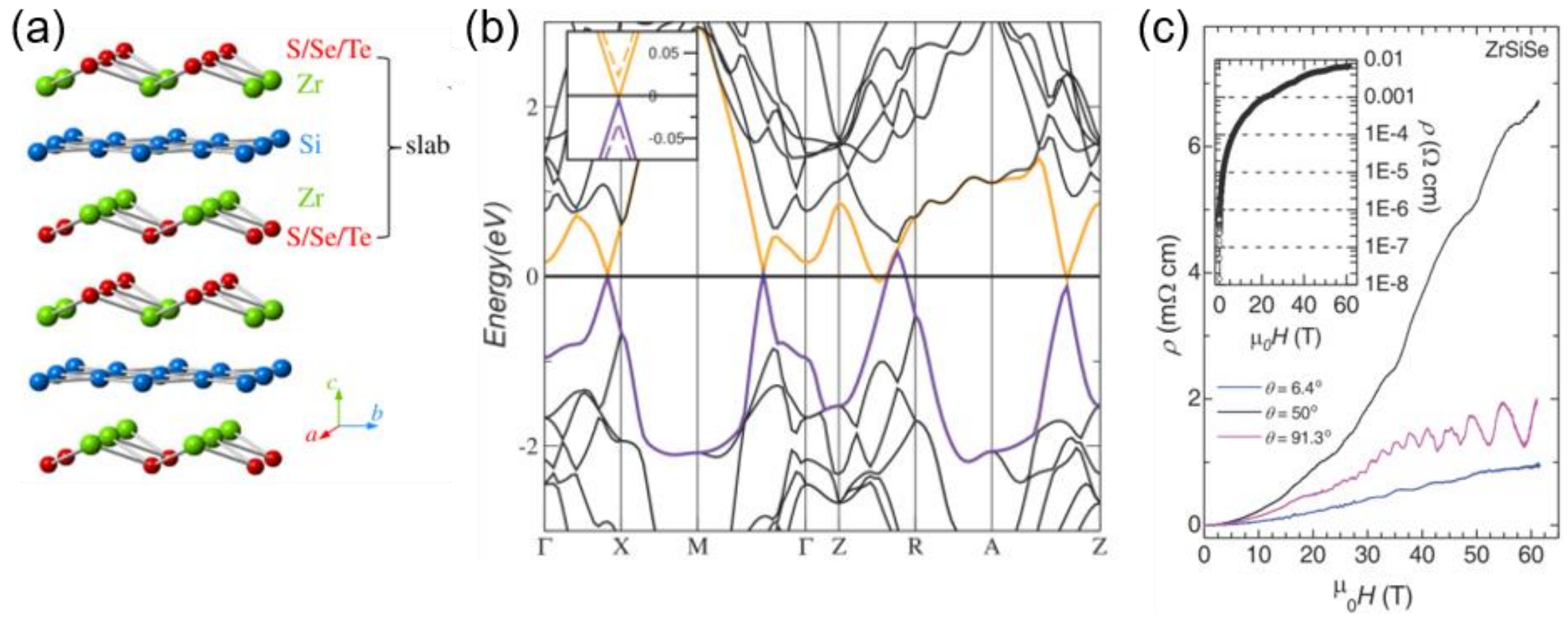

**Figure 12.** (a) Illustrated crystal structure of ZrSiSe. Adapted from Ref. [124]. (b) Electronic band structures for ZrSiSe. Inset: dashed SOC bands in comparison to solid non-SOC along ΓX line which highlights the effect of SOC. (c) Resistivity as a function of magnetic field for ZrSiSe. Adapted from Ref. [24].

Before ending this section, we draw a plot to display all the XMR materials marked with formulas and different symbols, as shown in figure 13. The *x* and *y* axis are measurement magnetic field and XMR value, respectively. If one material is tested under various (maximum) magnetic fields by different groups, the XMR taken at the highest field is adopted. Some outstanding materials deserve highlight here. For the magnetic field below 20 T that is commonly produced by a superconducting magnet, the maximum XMR comes from α-As, an elemental material. For the field between 20 T and 40 T that is usually from a resistive (water cooling) magnet or hybrid magnet, $PtBi_2$ yields the largest XMR. For the higher range of pulse magnetic field, $WP_2$ and ZrSiSe are in the top ranking. Since all the values are from experiments, they strongly depend on crystal quality, sample shape and size, and contact. Hence the positions in the plot are not absolute. Tables 1-3 detail the specific XMR values and measurement conditions.

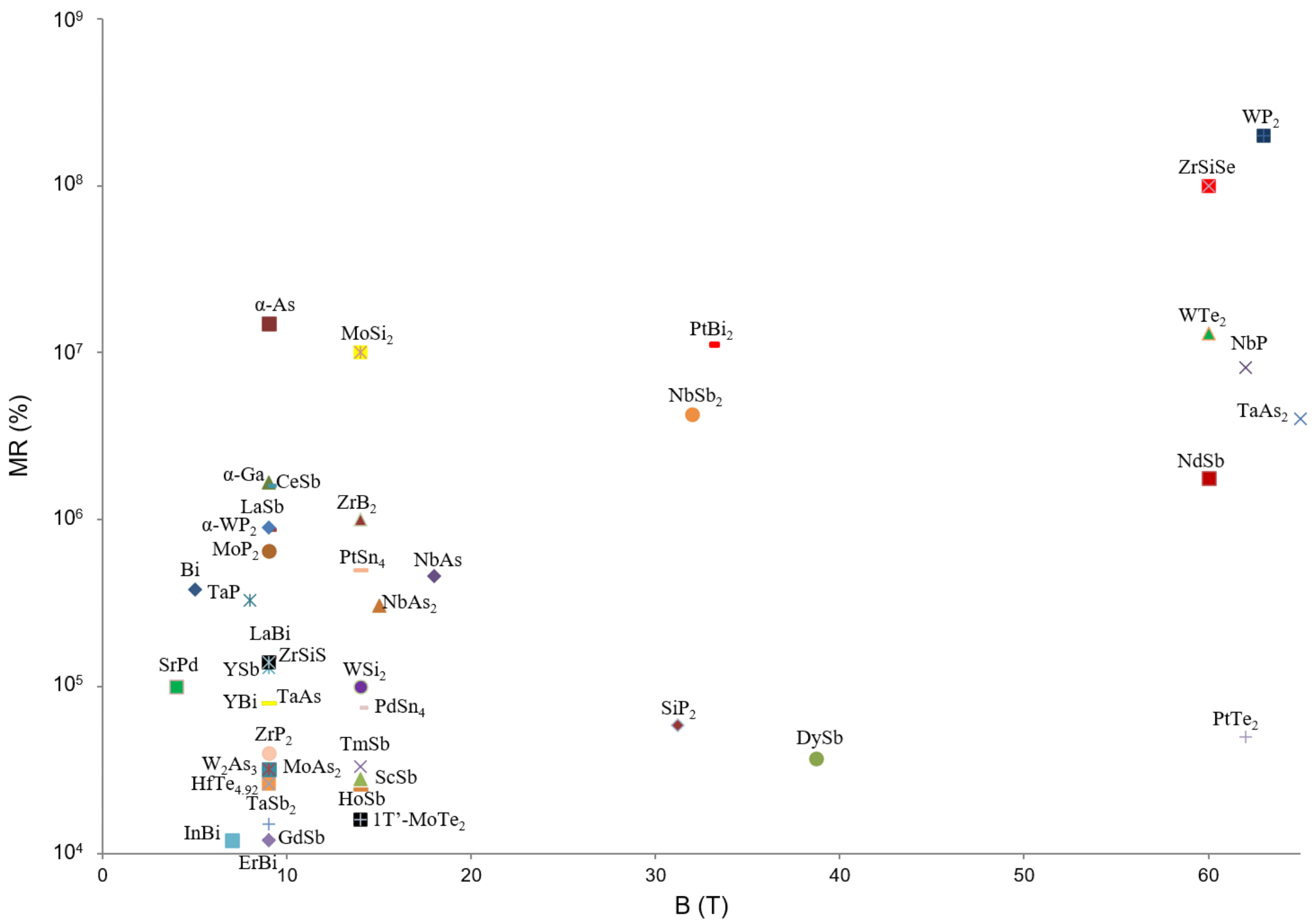

**Figure 13.** XMR of typical materials reported in past few years [2, 5-7, 11, 12, 14, 17, 18, 20, 21, 24, 28, 34, 36-40, 44-47, 50-53, 62, 64-67, 74, 76-78, 80, 86, 89-91, 94, 115]. The references are from Tables 1-3.

## 3. Proposed mechanisms for XMR

The ultimate task of physics research is to disclose the physical origins and universal laws (if any) underlying various experimental phenomena that provide guidance for the next step of research. For XMR, various potential mechanisms have been proposed, including electron-hole compensation, steep band, ultrahigh mobility, high RRR, topological fermions, etc. It turns out that some mechanisms play a leading role in certain systems, while more are far from clearly defined. Table 4 lists the proposed mechanisms and investigation methods for XMR materials.

**Table 4.** Proposed mechanisms and investigation methods for XMR materials.

| Material | Proposed mechanism(s) | Methods | Reference |
| --- | --- | --- | --- |
| α-As | compensation, high mobility | transport measurements, SdH oscillations | [6] |
| α-Ga | compensation | transport measurements, first-principles calculations, dHvA and SdH oscillations | [7] |
| CrP | compensation | transport measurements, DFT calculations and SdH oscillations | [33] |
| NbP | compensation, high mobility | transport measurements, *ab* initio calculations, SdH oscillations | [34] |
| TaP | compensation | transport measurements, SdH oscillations | [11, 125] |
| $WP_2$ | compensation, high mobility, RRR | transport measurements, ARPES, first-principles calculations, SdH oscillations | [21] |
| α-$WP_2$ | compensation, RRR | transport measurements, band structure calculations | [63] |
| $ZrP_2$ | compensation | transport measurements, ARPES, band structure calculations | [62] |
| TaAs | compensation | transport measurements, SdH oscillations, first-principles calculations | [12] |
| NbAs | compensation, high mobility | transport measurements | [35] |
| NbAs | cooperation of compensation and topological protection mechanism | transport measurements, SdH oscillations | [14] |
| $MoAs_2$ | compensation | transport measurements | [73] |
| $NbAs_2$ | compensation | transport measurements, SdH oscillations | [74] |
| $TaAs_2$ | compensation | transport measurements, SdH oscillations, band structure calculations | [75] |
| $Cd_3As_2$ | compensation | transport measurements | [126] |
| $W_2As_3$ | compensation, high mobility | transport measurements, SdH oscillations | [86] |
| GdSb | compensation, high mobility | transport measurements, first-principles calculations | [37] |
| HoSb | compensation, high mobility | transport measurements, SdH oscillations | [39] |
| LaSb | combination of compensation and a particular orbital texture on the electron pocket | transport measurements, band structure calculations, SdH oscillations | [41] |
| NdSb | particular orbital texture, compensation | transport measurements, SdH oscillations | [44] |
| ScSb | compensation | transport measurements, band structure calculations, SdH oscillations | [45] |
| TmSb | compensation, high mobility | transport measurements, ARPES, first-principles calculations, SdH oscillations | [46] |
| YSb | compensation | transport measurements, calculations, SdH oscillations | [47] |
| YSb | compensation, high mobility | transport measurements, ARPES, first-principles calculations, SdH oscillations | [48] |
| LaBi | compensation, high mobility | transport measurements, band structure calculations, SdH oscillations | [49] |
| YBi | compensation | transport measurements, first-principles calculations, SdH oscillations | [51] |
| ErBi | compensation, high mobility | transport measurements, band structure calculations | [52] |
| InBi | topological fermions (small $m^*$ and large carrier mobilities), compensation, RRR | transport measurements, first-principles calculations | [53] |
| $MoTe_2$ | compensation | transport measurements | [18] |
| $WTe_2$ | compensation, RRR | transport measurements | [58] |
| $WTe_2$ | compensation | transport measurements, SdH oscillations | [59] |
| $WTe_2$ | compensation | transport measurements, electronic structure calculations | [2] |
| $WTe_2$ | compensation, high mobility | transport measurements | [60] |
| $TbTe_3$ | compensation | transport measurements | [79] |
| $HfTe_{4.98}$<br>$HfTe_{4.92}$<br>$HfTe_{4.87}$ | compensation, high mobility | transport measurements | [67] |
| $TaSe_3$ | compensation | transport measurements, SdH oscillations | [92, 127] |
| $MoSi_2$ | combined effect of Zeeman effect-driven carrier compensation and ultrahigh carrier mobility | transport measurements, dHvA oscillations, electronic structure calculations | [65] |
| $WSi_2$ | compensation, high mobility | transport measurements, band structure calculation and dHvA oscillations | [66] |
| $PtTe_2$ | moderate degree of charge carrier compensation | transport measurements, band structure calculations, SdH oscillations | [57] |
| $PtBi_2$ | compensation | transport measurements, *ab* initio calculations, SdH oscillations | [80] |
| SrPd | compensation, high mobility | first-principles calculations | [94] |
| WC | compensation | transport measurements, dHvA oscillations | [95] |
| $ZrB_2$ | compensation, high mobility | transport measurements, band structure calculations, SdH oscillations | [113] |
| ZrSiSe | compensation | transport measurements, DFT calculations, dHvA oscillations | [24] |
| $\alpha$-$WP_2$ | high mobility | transport measurements, first-principles calculations, dHvA oscillations | [64] |
| $Cd_3As_2$ | high mobility | transport measurements, SdH oscillations | [84] |
| LaSb | high mobility, diminishing Hall effect | transport measurements, SdH oscillations, first-principles calculations | [42] |
| $NbSb_2$ | change of the Fermi surface induced by magnetic field which is related to the Dirac-like point, plus orbital MR expected for high mobility metals | transport measurements, first-principles calculations | [77] |
| $HfTe_5$ | high mobility | transport measurements, SdH oscillations | [70] |
| $PtSn_4$ | RRR | transport measurements, dHvA and SdH oscillations, band structure calculations | [90, 128] |
| $PtBi_{2-x}$ | RRR | transport measurements, dHvA and SdH oscillations | [81] |
| $MoAs_2$ | carrier motion on the Fermi surfaces with dominant open-orbit topology | transport measurements, first-principles calculations, ARPES | [28] |
| $Bi_2Te_3$ | zero-gap band structures with Dirac linear dispersion | transport measurements | [87] |
| $Cd_3As_2$ | lifting of topology protection by applied magnetic field | transport measurements, SdH oscillations | [85] |
| CeSb | semimetallic band structures | transport measurements, SdH oscillations | [38] |
| LaSb | metal-to-insulator-like transition | transport measurements, SdH oscillations | [40] |
| NdSb | time-reversal symmetry breaking | transport measurements, band structure calculations, | [43] |
| | within the antiferromagnetic state leads to the enhancement of backscattering | SdH oscillations, dHvA oscillations | |
| $HfTe_2$ | semimetallic nature with coexisting electrons and holes on the Fermi surface and the additional strong orbital mixing of Te *p* and Hf *d* states | transport measurements, *ab* initio calculations | [54] |
| $PdTe_2$ | well fitted according to the single-band model | transport measurements, dHvA oscillations | [19] |
| $PdSn_4$ | topological fermions, scaling property of MR that follows Kohler's rule over a wide magnetic field and temperature range | transport measurements, ARPES, dHvA oscillations | [89] |
| $SiP_2$ | topology of Fermi surface | transport measurements, dHvA oscillations, SdH oscillations, band structure calculations | [91] |
| $TaPdTe_5$ | unconventional quasiparticles | transport measurements, first-principles calculations, SdH oscillations, dHvA oscillations | [119] |

### *3.1 Compensation mechanism*

Compensation mechanism refers to two bands of charge carriers, i.e., electron-type carriers and hole-type carriers in metals and semiconductors. This is a strong assumption that there are only two bands participating transport process. This is impossible for most cases in which there are complex Fermi surfaces and pockets. Each pocket corresponds to an effective carrier mass, a carrier mobility and a carrier concentration or density, as well as a specific dependence on magnetic field and temperature. The actual transport process in real materials is complicated. However, the two-band model is a useful preliminary approximation, namely simplifying the carriers as two bands, one electron-type band and one hole-type band. The respective carrier density and mobility are $n_e$, $n_h$, $\mu_e$ and $\mu_h$.

According to the model, magnetic field dependence of longitudinal resistivity and Hall resistivity can be represented by

$$\rho_{\mathrm{xx}}(B)=\frac{(n_{\mathrm{e}}\mu_{\mathrm{e}}+n_{\mathrm{h}}\mu_{\mathrm{h}})+(n_{\mathrm{e}}\mu_{\mathrm{h}}+n_{\mathrm{h}}\mu_{\mathrm{e}})\mu_{\mathrm{e}}\mu_{\mathrm{h}}B^2}{e\left[(n_{\mathrm{e}}\mu_{\mathrm{e}}+n_{\mathrm{h}}\mu_{\mathrm{h}})^2+(n_{\mathrm{h}}-n_{\mathrm{e}})^2\mu_{\mathrm{e}}^2\mu_{\mathrm{h}}^2B^2\right]}, \quad (1)$$

$$\rho_{\mathrm{yx}}(B)=\frac{(n_{\mathrm{h}}\mu_{\mathrm{h}}^2-n_{\mathrm{e}}\mu_{\mathrm{e}}^2)B+(n_{\mathrm{h}}-n_{\mathrm{e}})\mu_{\mathrm{e}}^2\mu_{\mathrm{h}}^2B^3}{e\left[(n_{\mathrm{e}}\mu_{\mathrm{e}}+n_{\mathrm{h}}\mu_{\mathrm{h}})^2+(n_{\mathrm{h}}-n_{\mathrm{e}})^2\mu_{\mathrm{e}}^2\mu_{\mathrm{h}}^2B^2\right]}, \quad (2)$$

$$\mathrm{MR}(B)=\frac{(n_{\mathrm{e}}\mu_{\mathrm{e}}+n_{\mathrm{h}}\mu_{\mathrm{h}})^2+\mu_{\mathrm{e}}\mu_{\mathrm{h}}(n_{\mathrm{e}}\mu_{\mathrm{e}}+n_{\mathrm{h}}\mu_{\mathrm{h}})(n_{\mathrm{h}}\mu_{\mathrm{e}}+n_{\mathrm{e}}\mu_{\mathrm{h}})B^2}{(n_{\mathrm{e}}\mu_{\mathrm{e}}+n_{\mathrm{h}}\mu_{\mathrm{h}})^2+(n_{\mathrm{h}}-n_{\mathrm{e}})^2\mu_{\mathrm{e}}^2\mu_{\mathrm{h}}^2B^2}-1. \quad (3)$$

When $n_{\mathrm{e}}=n_{\mathrm{h}}$, i.e., the perfect compensation, we have

$$\mathrm{MR}(B)=\mu_{\mathrm{e}}\mu_{\mathrm{h}}B^2, \quad (4)$$

$$\rho_{\mathrm{yx}}(B)=\frac{(n_{\mathrm{h}}\mu_{\mathrm{h}}^2-n_{\mathrm{e}}\mu_{\mathrm{e}}^2)B}{e(n_{\mathrm{e}}\mu_{\mathrm{e}}+n_{\mathrm{h}}\mu_{\mathrm{h}})^2}. \quad (5)$$

Under this condition, the MR is proportional to $B^2$ and the Hall resistivity is proportional to $B$ [129]. More importantly, the maximum of MR occurs when $n_{\mathrm{e}}=n_{\mathrm{h}}$, which is the compensation mechanism for XMR. From the above description, it is easy to know that the compensation mechanism applies to highly simplified systems. Inspired by this, many groups have conducted elaborate experiments to verify or challenge the compensation mechanism.

As a representative XMR material, $WTe_2$ shows an unsaturated XMR of $1.3 \times 10^7\%$ at 0.53 K and 60 T [2], which leads to systemic investigations into its underlying mechanism. Thus, we take $WTe_2$ as an example to discuss the compensation mechanism. Luo et al. systematically measured the Hall effect of $WTe_2$ at various temperatures [129]. By carefully fitting the Hall resistivity to the two-band model (figures 14(a) and 14(b)), the temperature dependence of carrier density and mobility for both electron- and hole-type carriers were determined. They observed a sudden increase in the hole density below ~160 K (figure 14(d)), which is likely associated with the temperature-induced Lifshitz transition reported by a previous photoemission study [130]. In addition, a more pronounced reduction in electron density occurs below 50 K (figure 14(d)), giving rise to comparable electron and hole densities at low temperatures. These observations indicate a possible electronic structure change below 50 K, which is claimed as the direct driving force of the electron-hole compensation and XMR as well, although numerical simulations imply that this material is unlikely to be a perfectly compensated system.

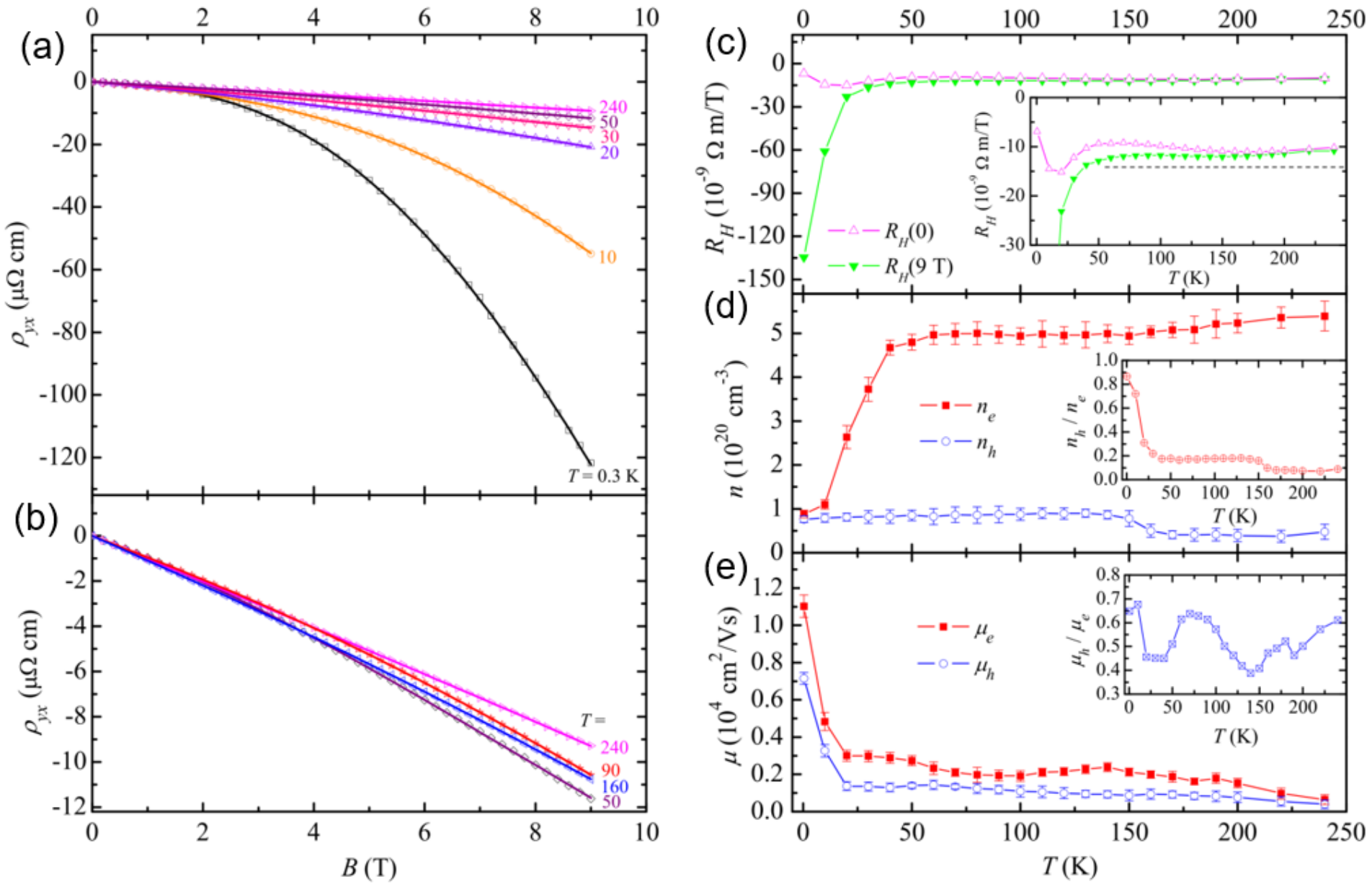

**Figure 14.** (a)(b) Field dependence of Hall resistivity at various temperatures. The curve for 0.3 K is the non-oscillating background <$\rho_{yx}$>. The open symbols are experimental data and the solid lines are numerical fitting to Eq. 2. (c) Hall coefficient $R_H$ as a function of $T$. The solid symbols represent $R_H$ defined by $\rho_{yx}/B$ at $B$ = 9 T, and the open symbols stand for $R_H$ determined from the initial slope of $\rho_{yx}(B)$ at $B$ approaching 0. The dashed line in the inset is a guide line to eyes. (d) and (e) display the temperature dependent carrier density and mobility, respectively. The insets in (d) and (e) show the plot of $n_h/n_e$ and $\mu_h/\mu_e$ versus $T$. Adapted from Ref. [129].

Further consideration is focused on tuning the ratio of carrier densities, namely changing the situation of compensation, to check the effect on XMR. One method is to introduce different degrees of deficiency during the process of sample growth. Ali et al. improved the growth technique, synthesizing high-quality single crystals of $WTe_2$ using a Te flux followed by a cleaning step involving self-vapor transport. The method yields consistently higher-quality single crystals than are typically obtained via halide-assisted vapor transport methods. It is generally believed that high-quality crystals possess better compensation in carrier densities. This is confirmed by another work in the reversal way, i.e., introducing more deficiency of Te by annealing the as-grown crystals in vacuum [60]. Gong et al. compared the naturally grown and annealed samples and found the Te vacancies, understood in terms of electron doping, indeed shift the balance of electron and hole carrier densities and suppress the XMR.

Another effort is to directly tune the location of Fermi level that critically affects the electron-hole balance, via applying a gate to a thin film or flake device of $WTe_2$. Wang et al. conducted an in situ tuning experiment on a thin $WTe_2$ film with an electrostatic doping approach (figure 15(a)) [59]. They observed a nonmonotonic gate dependence of the MR (figure 15(b)). The MR reaches a maximum (10600%) in thin $WTe_2$ film at certain gate voltage where electron and hole concentrations are balanced, indicating that the charge compensation is the dominant mechanism of the XMR. Besides, the temperature-dependent MR exhibits similar tendency with the carrier mobility when the charge compensation is retained, revealing that distinct scattering mechanisms may be at play for the temperature dependence of magneto-transport properties. This work experimentally confirms that the maximum of XMR is achieved at the compensation point with other conditions set. Nevertheless, it does not rule out other potential factors or mechanisms that also contribute to the XMR.

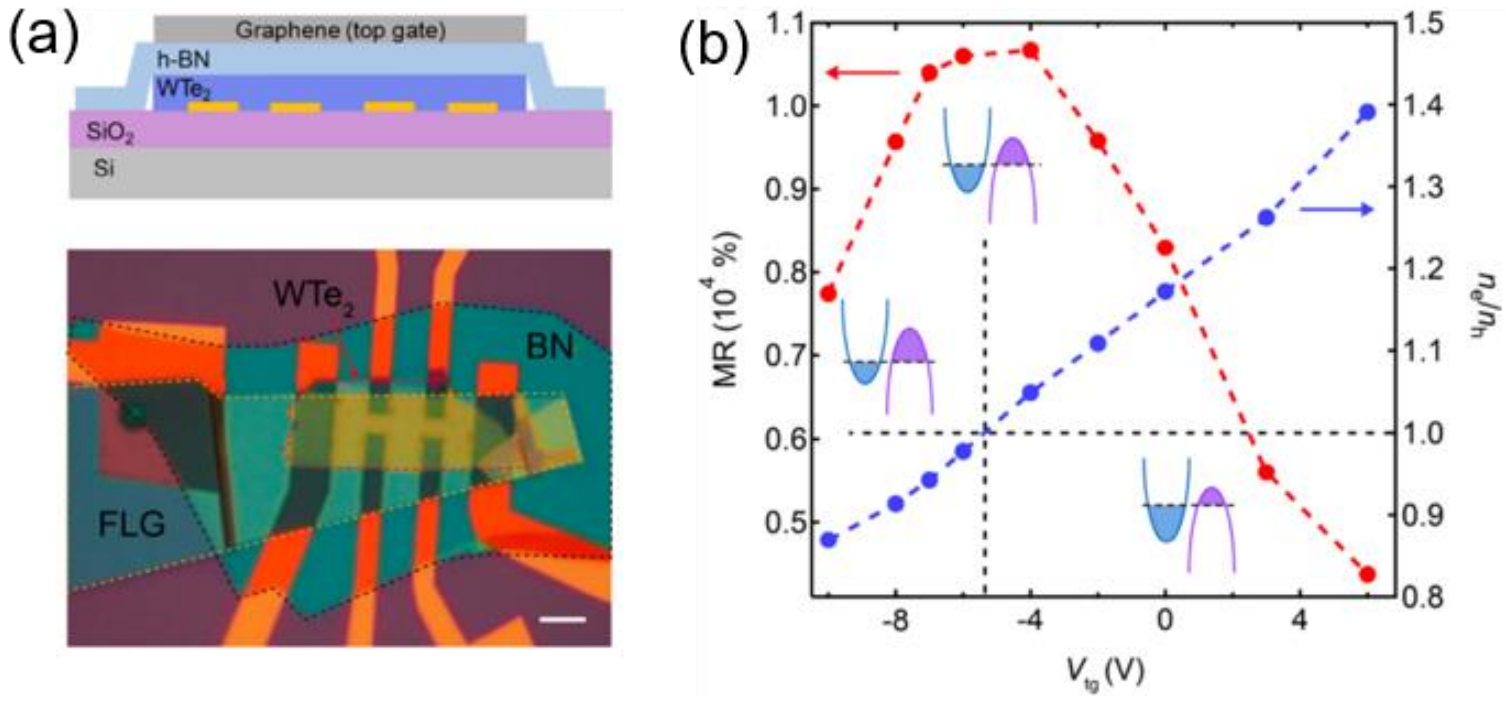

**Figure 15.** (a) Optical image of 10 nm thick $WTe_2$ device; the boundary of the $WTe_2$ film is indicated by the red dashed line. The edges of few-layer graphene and *h*-BN are indicated by the yellow and black dashed lines, respectively. The length of the white scale bar is 5 μm. The upper panel is the cross-sectional schematic structure of the device. (b) Comparison of the gate voltage-dependent MR and carrier densities ratio. The MR curve reaches the maximum by tuning the carrier concentration through the charge compensation point ($n_e/n_h = 1$). The three schematic bands represent different scenarios for charge distribution in electron (blue) and hole (purple) pockets at different gate voltage ranges. Adapted from Ref. [59].

ARPES is another useful tool to directly estimate the carrier density of various Fermi pockets. As seen in figure 16(a), the calculated band structures show that the hole pocket and the electron pocket are of almost the same size [131]. This is confirmed by the ARPES results [132]. The ARPES intensity maps shown in figure 16(b) represent constant energy contours of the states at the Fermi level. The states appear as two distinct pockets. One is determined to be an electronlike (e) pocket, and the other a holelike (h) one. The size of the pockets at the Fermi level is almost exactly the same, 0.018 Å$^{-2}$. Taking into account spin degeneracy and the fact that there are two such pockets in the surface Brillouin zone, the carrier concentration in the electron pockets is about $1.8 \times 10^{13}$ cm$^{-2}$ and is nearly perfectly compensated by the same concentration of holes. This suggests that carrier compensation should be considered the primary source of the XMR in $WTe_2$. More research efforts using ARPES can be found in Refs.[133, 134].

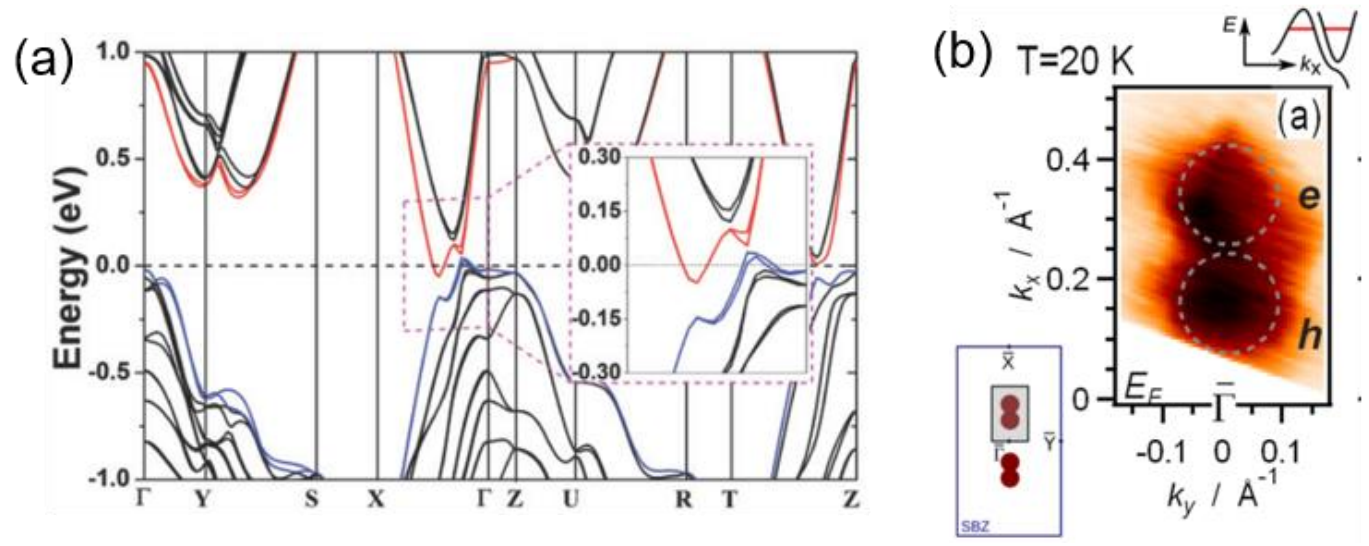

**Figure 16.** (a) Calculated band structures of $WTe_2$ along high symmetry directions. Inset: enlarged view near band inversion. Adapted from Ref. [131]. (b) Constant energy contours at 20 K in the region of the Brillouin zone where the bands of $WTe_2$ cross the Fermi level. Electronlike and holelike pockets are marked with e and h, respectively. Adapted from Ref. [132].

Although the electron-hole compensation is often considered to be the main mechanism of XMR (see Table 4), it has some limitations. First, as we already know, compensation model is based on a highly simplified case which discards too many necessary features and details in actual transport process. This usually brings difficulty when we talk about complex cases, e.g., multiband systems ($\geq$ 3), especially with distinctly different effective masses and mobilities. Second, compensation is a necessary condition for the XMR to reach its maximum with all other conditions set unchanged. However, it is not a necessary condition for the occurrence of XMR. When the compensation balance is broken, the XMR can still survive, though the magnitude of XMR may decrease. Third, related with the second point, another important characteristic of XMR, i.e., unsaturation up to very high magnetic fields, can still hold even for the situation diverging from compensation. This suggests that the compensation mechanism is unlikely to account for the (only) origin of XMR. More factors should be considered and carefully examined.

### *3.2 High mobility*

The carrier mobility is defined as $\mu = e\tau/m^*$ and the cyclotron frequency as $\omega_\mathrm{c} = eB/m^*$, where $\tau$ is the relaxation time. Hence, we have $\omega_\mathrm{c}\tau = \mu B$, both of which are dimensionless. In the two-band model, the MR is derived as $\mu_\mathrm{e}\mu_\mathrm{h}B^2$ for the perfect compensation, which indicates that the MR is proportional to $\mu_\mathrm{e}\mu_\mathrm{h}$. A. B. Pippard has demonstrated that the MR shows varying dependence for different scales of $\omega_c\tau$, i.e., $\mu B$, in real metals [135]. Taking copper as an example, the MR with $B$ along the [100] direction follows a quadratic dependence in the low limit of $\mu B$, and tends to saturate in the high limit of $\mu B$, and runs linearly in the intermediate region (figure 17). The proportion of electrons and holes, orbit shapes and variations of $\tau$ all take a role in the evolution process. From the relation of MR and mobility, it is easy to know MR is positively correlated with mobility. That is, a higher mobility always leads to a larger MR. One noteworthy point is that the MR of copper tends to saturate at higher values of $\omega_c\tau$. This may be due to possible changes of Fermi surfaces upon increasing magnetic field. Though copper is not an XMR material, the dependence on $\mu B$ is generally applicable.

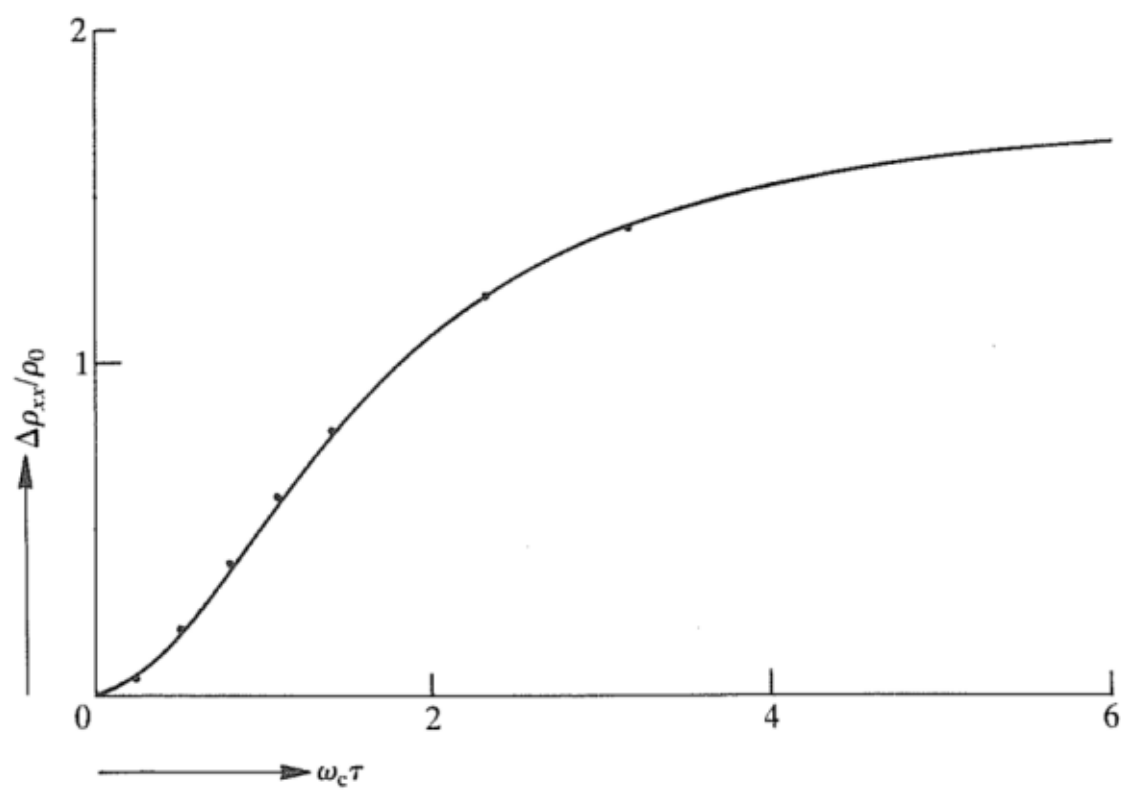

**Figure 17.** Computation of simplified model for copper with $B$ along [100], comparing with points taken from the measurements of Ref. [136]. Adapted from Ref. [135].

The role of mobility is revealed by specifically designed experiments. Lv et al. observed an XMR of 8.74 × $10^5$% at 2 K and 9 T for α-$WP_2$ single crystal, in which the electron-hole compensation was determined to maintain in a large temperature range (2-100 K). However, when the temperature rises from 2 K to 100 K, the XMR undergoes a dramatic decrease. Meanwhile, the mobility also decreases by two orders of magnitude, which indicates that the XMR strongly depends on the high mobility ($\mu_h$ = 5.36 × $10^4$ $cm^2$/V s, $\mu_e$ = 2.04 × $10^5$ $cm^2$/V s) of carriers rather than the electron-hole compensation [64]. It is generally found that the XMR materials mostly have an ultrahigh mobility. For example, the average mobility of ErBi is 1.27 × $10^4$ $cm^2$/V s [52] and the carrier mobility of TaAs is 1.8 × $10^5$ $cm^2$/V s [12]. The electron mobility of LaSb and YSb is 6.73 × $10^3$ $cm^2$/V s [42], and 4.02 × $10^4$ $cm^2$/V s [48], respectively.

Another aspect to review the effect of mobility is from the relation of $\mu = e\tau/m^*$, in which $\mu$ is inversely proportional to $m^*$. As we know, Fermi surfaces with a small area result in carriers of small effective mass. This explains why most XMR materials are semimetals that have small pockets and minor DOS at the Fermi level. Through quantum oscillation measurements and analyses, information of Fermi surfaces and corresponding carriers, including the extremal cross-section areas of Fermi pockets, carrier concentration and effective carrier mass, can be retrieved. For instance, a small effective electron mass of $m^*$ = 0.03 $m_e$ and a high mobility $\mu$ = 4.6 × $10^4$ $cm^2$/V s are obtained for $ZrTe_5$ [109]. For NbP, the SdH oscillation analyses give a small effective mass ($m^*$ = 0.076 $m_e$) and an ultrahigh mobility (5 × $10^6$ $cm^2$/V s) that contribute to a significant XMR up to 8.1 × $10^6$% [34]. Similarly, the small effective electron mass $m_\alpha^*$ (0.066 $m_e$) and effective hole mass $m_\beta^*$ (0.033 $m_e$) accompany high mobilities ($\mu_\alpha$ = 1.9 × $10^5$ $cm^2$/V s, $\mu_\beta$ = 1.9 × $10^6$ $cm^2$/V s) in the XMR material NbAs [14].

### *3.3 RRR*

XMR is also found to rely on crystal quality of samples which is usually represented by the residual resistivity ratio, i.e., RRR. RRR is defined as the ratio of the resistivity of a material at room temperature and at 0 K. Since 0 K can never be reached in practice, a low temperature near 0 K is usually used. Depending on the amount of impurities and other crystal defects, RRR may vary greatly for a single material, so it can serve as a rough indicator of sample purity and overall quality. Because the resistivity generally increases with the increasing defects, a large RRR is associated with pure samples.

As is introduced in subsection 3.1, Ali et al. have devoted to high quality growths of single crystals of $WTe_2$ using a Te flux followed by a cleaning step involving self-vapor transport, which yields consistently higher quality single crystals than are typically obtained via halide-assisted vapor transport methods [27]. The RRR and MR of the crystals are checked and compared. Figure 18 shows the correlation of the MR at 2 K and 9 T and the average carrier mobility (as extracted from Lorentz fitting) with the RRR value for various crystals fabricated by different methods. For the flux method, three different cooling rates were used to control the RRR value and, correspondingly, the MR and mobility. The highest RRR was obtained with the slowest-cooling rate. The correlation of $\mu_{\rm avg}$ with RRR appears roughly linear while the MR appears to follow a roughly square law correlation, consistent with the Lorentz MR law. This study, together with another work by Gong et al. [60] , provides direct evidence for the positive effect of RRR on XMR, and again demonstrates the effect of carrier mobility. Besides, similar behaviors and rules have been confirmed in other systems, i.e., $PtBi_{2-x}$ [81] and InBi [53].

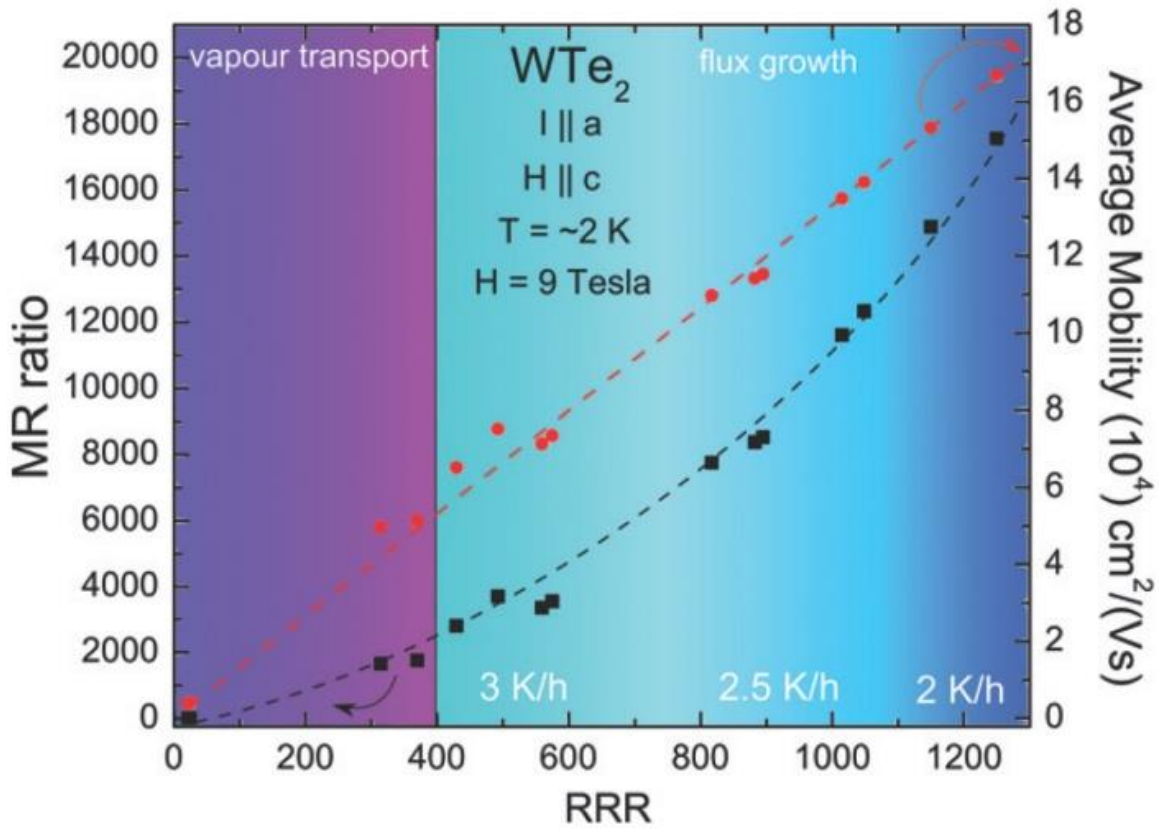

**Figure 18.** MR and average carrier mobility versus RRR. MR follows the left axis (black squares) while $\mu_{\rm avg}$ follows the right axis (red circles). The dashed lines are guides to the eye with the black line meant to show the non-linear correlation of the MR with RRR and the red line meant to show the linear correlation of $\mu_{\rm avg}$ with RRR. Shading is used to distinguish crystals made using the flux method at particular cooling rates from crystals made using vapor transport. Adapted from Ref. [27].

In spite of the success of explaining XMR by RRR, the scenario has its ambiguity. RRR is a simple and rough index, behind which are specific parameters that affect transport and XMR, including electronic structure (especially Fermi surface), carrier mobility, effective mass and relaxation time, etc. Changes in the amount of impurities and defects will cause changes in the Fermi surface, which is manifested in the shape of Fermi surface and the ratio of electron and hole carriers. The scattering by impurities and defects will also change, which will affect the carrier mobility and effective mass. The main factors influencing XMR are the ratio of carriers (through compensation mechanism) and carrier mobility [60]. Although carrier mobility generally benefits from the improvement of crystal quality, the ratio of carriers may not always be improved. Note that the perfect compensation does not necessarily occur at the stoichiometric point.

### 3.4 Steep band

Steep bands are in reference to linear or nearly liner energy dispersion, regardless of whether there is band crossing or inversion. For the parabolic dispersion in a 2D system as shown in figure 19(a), the energy dispersion relation is $E = \frac{\hbar^2 k^2}{2m^*} + \varepsilon_0$, where $\hbar$ is the reduced Planck constant, $k$ is the momentum and $\varepsilon_0$ is the energy of band vertex. The effective mass is thus derived as $m^* \propto 1/\frac{\partial^2 E}{\partial k^2}$, which is a finite value for the parabolic dispersion. However, for the linear dispersion band, zero effective mass is induced (figure 19(b)). According to the relation $\mu = e\tau/m^*$, the massless linear band means an ultrahigh mobility that is in favor of XMR. In real metals, the normal parabolic dispersion is related with trivial electrons, and the steep bands usually point to special or nontrivial electrons.

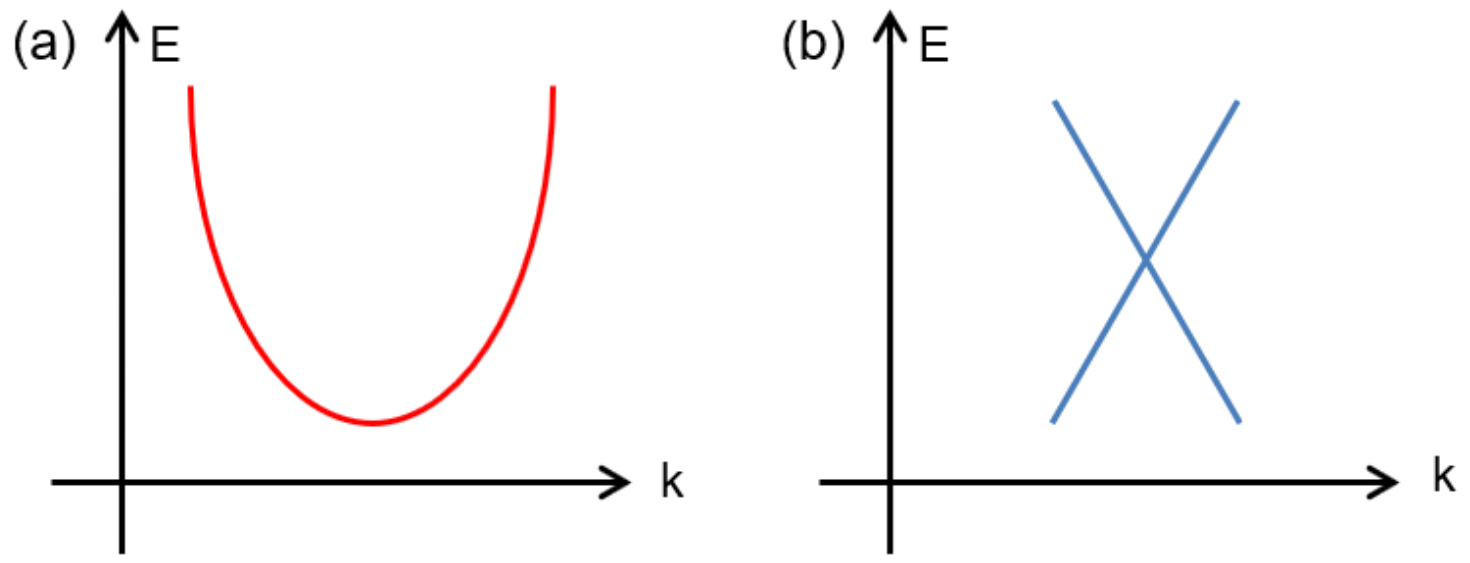

**Figure 19.** Schematic of (a) parabolic dispersion band and (b) linear dispersion band.

Liang et al. studied the XMR of DySb from a basis of electronic structure by SdH oscillation measurements and first-principles calculations [36]. Three possible origins of XMR are examined. Although a band inversion is found theoretically, suggesting that DySb might be topologically nontrivial, the inversion is deeply underneath the Fermi level, which rules out a topological nature of the XMR. The total densities of electron-like and hole-like carriers are not fully compensated, determined by the SdH oscillation analyses, showing that compensation is unlikely to account for the XMR. The XMR is eventually understood in terms of high mobility that is associated with the steep linear bands. Indeed, small effective masses ranging from 0.21 to 0.49 $m_e$ are revealed by the calculations, and a high carrier mobility $\mu_{avg}$ = 4800 $cm^2$/V s is retrieved from the Lorentz law fit.

### *3.5 Topological fermions*

The relationship between XMR and topological mechanism in topological materials is mainly reflected in two aspects. First, the electronic feature of topological materials is that the conduction band and the valence band cross at one point in the momentum space and the energy band is linear or nearly linear. It follows that the linear dispersion results in low-energy quasiparticles with zero (or very small) mass and high mobility at the intersection point. These characteristics are all promotive force of XMR. Second, for topological systems, the carriers' spin and momentum in the topological surface state are locked, and a conductive channel without backscattering is formed on the surface or boundary of the material, which improves the carrier mobility. For instance, Shekhar et al. determined an ultrahigh carrier mobility ~ $5 \times 10^6$ $cm^2$/V s, accompanied by a very small effective mass ~ 0.076 $m_e$ and a high Fermi velocity ~ $4.8 \times 10^5$ m $s^{-1}$, which is associated with the linear energy bands on the Fermi surface of Weyl semimetal NbP. Correspondingly, an extraordinary XMR phenomenon was discovered in NbP, i.e., up to $8.5 \times 10^5$% at 1.85 K and 9 T, $3.6 \times 10^6$% at 1.3 K and 30 T, and finally $8.1 \times 10^6$% at 1.5 K and 62 T, still without signature of saturation [34]. Similarly, for NbAs, an XMR of $4.62 \times 10^5$% was observed at 1.9 K and 18 T. Based on the analysis of SdH oscillations and other transport data, the XMR is believed to originate from electron-hole compensation and topological protection mechanism that strongly suppresses backward scattering and induces a high carrier mobility ($\mu_h = 1.9 \times 10^6$ $cm^2$/V s at 2 K) [14]. In addition, Okawa et al. unveiled the "hidden" 3D Dirac bands induced by spin-orbit interactions in a nonmagnetic semimetal InBi. The small $m^*$ and high mobility ($1.6 \times 10^4$ $cm^2/V$ s at 1.8 K) give rise to an unsaturated XMR, with the help of electron-hole compensation [53].

### 3.6 Other mechanisms

Other than these relatively more discussed mechanisms, there are some unusual mechanisms that can generate XMR or work together with one or more of the mechanisms listed above. Mangelsen et al. reported an experimental and theoretical study on the layered TMD material $HfTe_2$ that shows a large MR of 1350% at 2 K and 9 T in the absence of Dirac or Weyl points [54]. This result clearly distinguishes $HfTe_2$ from TMDs like $MoTe_2$ or $WTe_2$ which both exhibit larger MR and are viewed as Weyl semimetals. For the origin of the large MR, the simple picture of just one type of hole and electron compensation cannot be supported by the calculated band structure as multiple hole and electron pockets are predicted. Therefore, it is ascribed to carrier compensation and the mixed orbital character of the hole pockets.

Moreover, using ARPES combined with first-principles calculations and magnetotransport measurements, Lou et al. performed a comprehensive investigation on $MoAs_2$, which is isostructural to the $TmPn_2$ family and also exhibits quadratic XMR. Intriguingly, they found the unambiguously observed Fermi surfaces are dominated by an open-orbit topology extending along both the [100] and [001] directions in the 3D Brillouin zone. They further revealed the trivial topological nature of $MoAs_2$ by bulk parity analysis. Based on these results, they examined the proposed XMR mechanisms in other semimetals, and conclusively ascribed the origin of quadratic XMR in $MoAs_2$ to the carrier motion on the Fermi surfaces with dominant open-orbit topology [28].

To summarize this section, MR is a fundamental physical parameter to describe transport properties. The transport process in real metals involve a great many of variables, which makes the study of MR mechanisms really difficult. This is further complicated by numerous types of materials and strong crystallographic anisotropy. Nevertheless, it is variation and complexity that contribute one main interest of materials research and condensed matter physics. As shown in figure 20, for single crystals of tin with $B$ in different directions, while the curve taken under condition A saturates with MR ~ 6, the curve taken for condition B (magnetic field rotated by 30°) is still rising sharply at 75000%, which is indicative of highly anisotropic Fermi surfaces of tin. On the other hand, people always attempt to find unified and universal laws underneath experimental phenomena. It seems that compensation is the primary driving force of MR in most cases. Once compensation is established, a large MR can be expected. This does not mean that perfect compensation is necessary. For cases with poor compensation (sometimes well diverging from compensation), XMR can still exist. The factor that comes in the second is high mobility which always promotes a large MR. We may be aware of the fact that some of the mechanisms discussed above are more or less related to carrier mobility, including RRR, steep band and topological protection. Therefore, when we review the origins of new XMR materials, compensation and mobility are the first two factors to consider. In-depth research involves the structure of Fermi surfaces and the carrier properties that differ for materials. The last noteworthy point is the evolution of Fermi surfaces with the increase of magnetic field, which undoubtedly determines the nature and magnitude of XMR. A typical example is bismuth whose Fermi surface has recently been found to be very fragile and changeable upon increasing magnetic field [137].

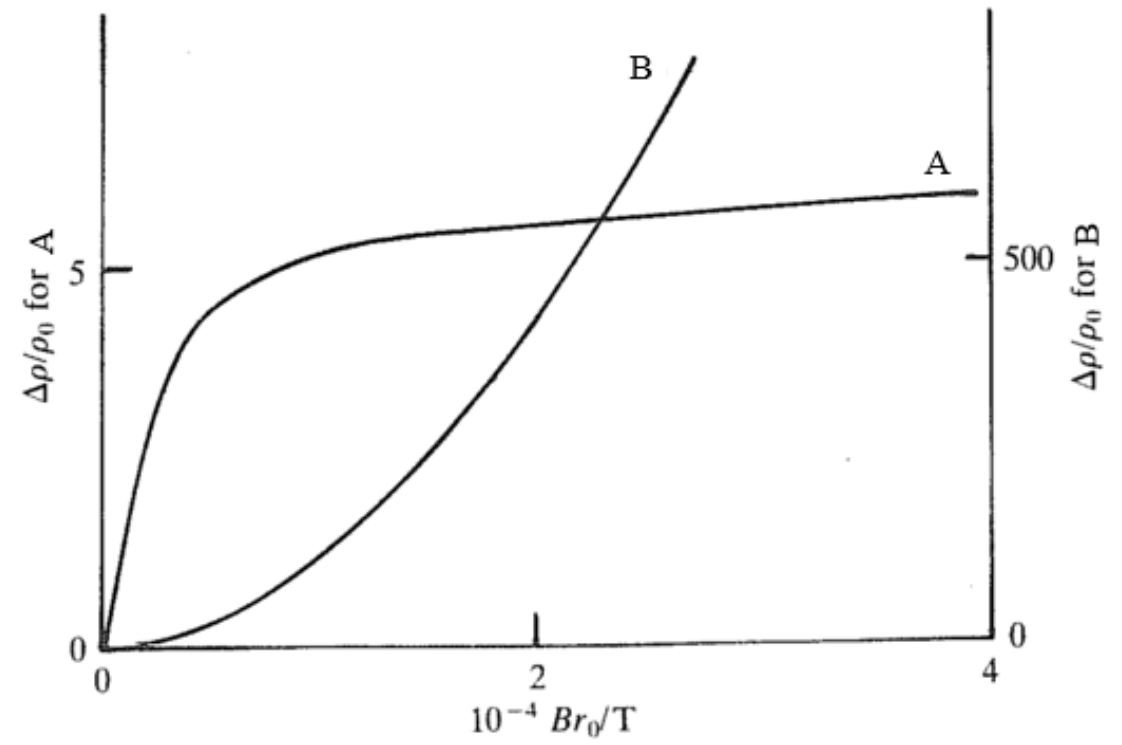

**Figure 20.** Transverse MR at 4 K of a single crystal of tin ($r_0$ = 11000) with current along the *c* axis and two different orientations of *B* in the basal plane. Adapted from Ref. [135].

## 4. Correlation with other systems

In the past two decades, layered and 2D materials and topological matters have been two mainstreams in condensed matter physics. The former started from the exfoliation and research of monolayer graphite, i.e., graphene [138], and the latter was triggered by the exploration of 2D (and 3D) quantum spin Hall insulators, i.e., topological insulators [139-141]. As shown in figure 21, the objective materials studied in these two directions are both largely overlapped or closely correlated with the XMR materials reviewed in this paper. Indeed, XMR was first proposed in the research of 2D materials and topological materials, and it has quickly grown into a large family and become a hot topic. This fact makes it even more important to elucidate the mechanism of XMR. On the one hand, as a fundamental parameter, XMR reflects the basic features of transport and the key information of Fermi surface. The research of XMR can definitely help understand the origin of nontrivial properties in topological materials and 2D materials. On the other hand, the disclosed novel properties in the interdisciplinary fields will lay a solid foundation for the design and development of functional devices, which will greatly expand the application scenarios of these materials. In this section, we will focus on discussing the characteristics of 2D materials and topological materials, namely layer structure related features and topological nontrivial properties respectively, as well as their relationship with XMR.

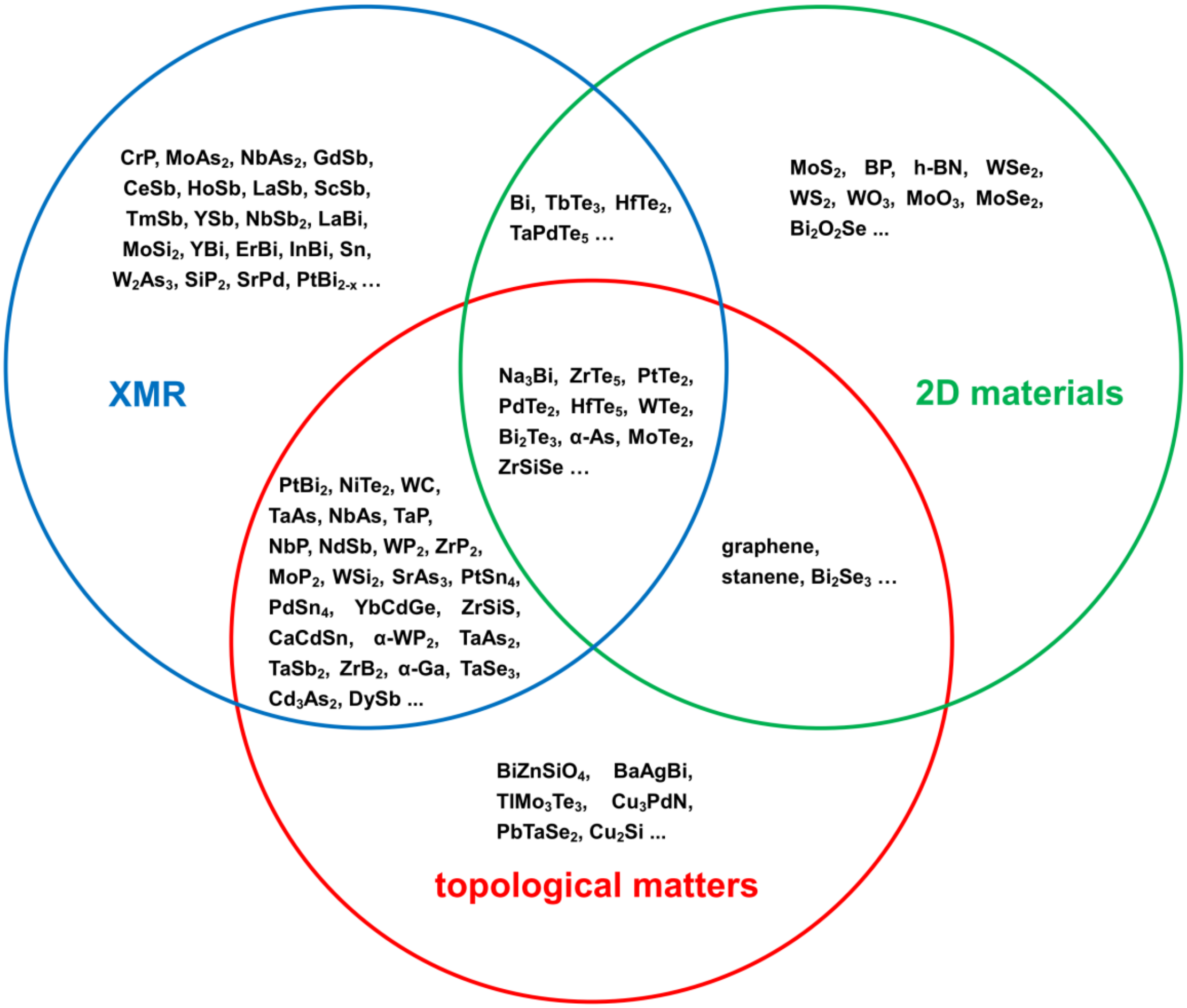

**Figure 21.** Overlap of XMR materials with 2D materials and topological matters.

### *4.1 2D materials*

One advantage of 2D materials is the possibility to fabricate miniaturized devices. For 2D materials, carrier migration is limited to the 2D plane and suppressed between planes, which induces many unique transport properties. The electronic structure largely depends on the thickness or layers of the material, which further modulates and enriches the transport behaviors. $WTe_2$ and $MoTe_2$ are taken as representative examples to describe these characteristics.

Figure 22(a) shows a schematic diagram of the crystal structure of $WTe_2$, which has an obvious layered structure. Liu et al. used scotch tape to exfoliate bulk $WTe_2$ into nanosheets, for which the atomic force microscopy images are shown in figures 22(b)-(d). It is found that a few layers of $WTe_2$ nanosheets are unstable in air, and four layers of $WTe_2$ nanosheets can only remain stable in air for 62 minutes. Figures 22(e)-(g) show the height profiles along the corresponding directions in figures 22(b)-(d). The thickness of monolayer $WTe_2$ is estimated to be 0.79 nm. In addition, defects in a few layers of $WTe_2$ nanosheets can induce a transition from metal to insulator, accompanied by the evolution of MR linetype from parabolic to linear and then to weak anti-localization [142]. As the thickness decreases, it is found that the MR of $WTe_2$ decreases. The XMR measured for bulk $WTe_2$ is as high as $1.3 \times 10^7$% at 60 T, as shown in figure 22(h) [2], while the $WTe_2$ thin film device (about 10 nm) fabricated by dry transfer technique just exhibits an XMR of $1.06 \times 10^4$% (figure 22(i)) [59]. A similar phenomenon was also confirmed in the experiment of Zhao et al. [143]. Das et al. further proposed that the balance state of electron-hole carriers in $WTe_2$ can be found only when there are at least three Te-W-Te layers, through first-principles calculations and high-resolution surface topography [144].

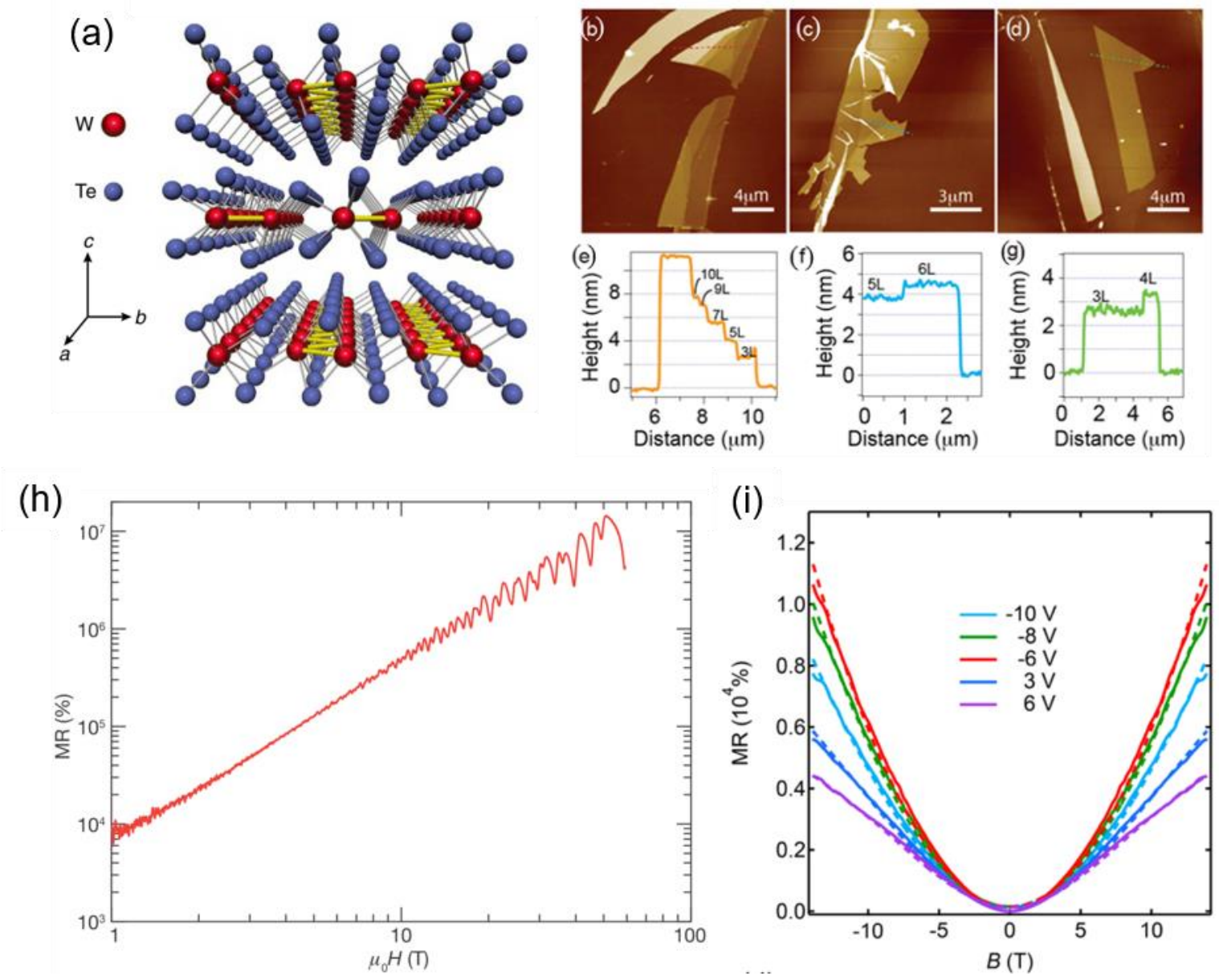

**Figure 22.** (a) Illustrated crystal structure of $WTe_2$ showing a layered structure. Adapted from Ref. [145]. (b)-(d) Atomic force microscopy images of typical $WTe_2$ flakes with different thickness. Height profiles along corresponding dashed lines are given in (e)-(g). Adapted from Ref. [142]. (h) XMR of $WTe_2$ up to 60 T at 0.53 K, with *I* parallel to *a* and *B* parallel to *c*. Adapted from Ref. [2]. (i) MR of thin film device. Adapted from Ref. [59].

$MoTe_2$ is another TMD with a significant XMR. Figure 23(a) shows the crystal structure of 1T'-$MoTe_2$, a monoclinic phase. Interestingly, the properties of $MoTe_2$ vary greatly between phases. Due to the weak vdW force between layers, it is easy to peel off into nanosheets. With the decrease in the number of layers, 2H-$MoTe_2$ possesses a gradually decreased band gap, and eventually becomes a direct band gap semiconductor for the single layer (figure 23(b)) [146]. While 1T'-$MoTe_2$ exhibits a significant XMR up to $1.6 \times 10^4$% at 1.8 K and 14 T, with distinct SdH oscillations in high fields, 2H-$MoTe_2$ does not show an XMR (figure 23(c)), which may be related to their different electronic structures. 1T'-$MoTe_2$ has been proven to be a semimetal with a maximum carrier mobility of 4000 $cm^2$/V s, whereas 2H-$MoTe_2$ is a semiconductor [18].

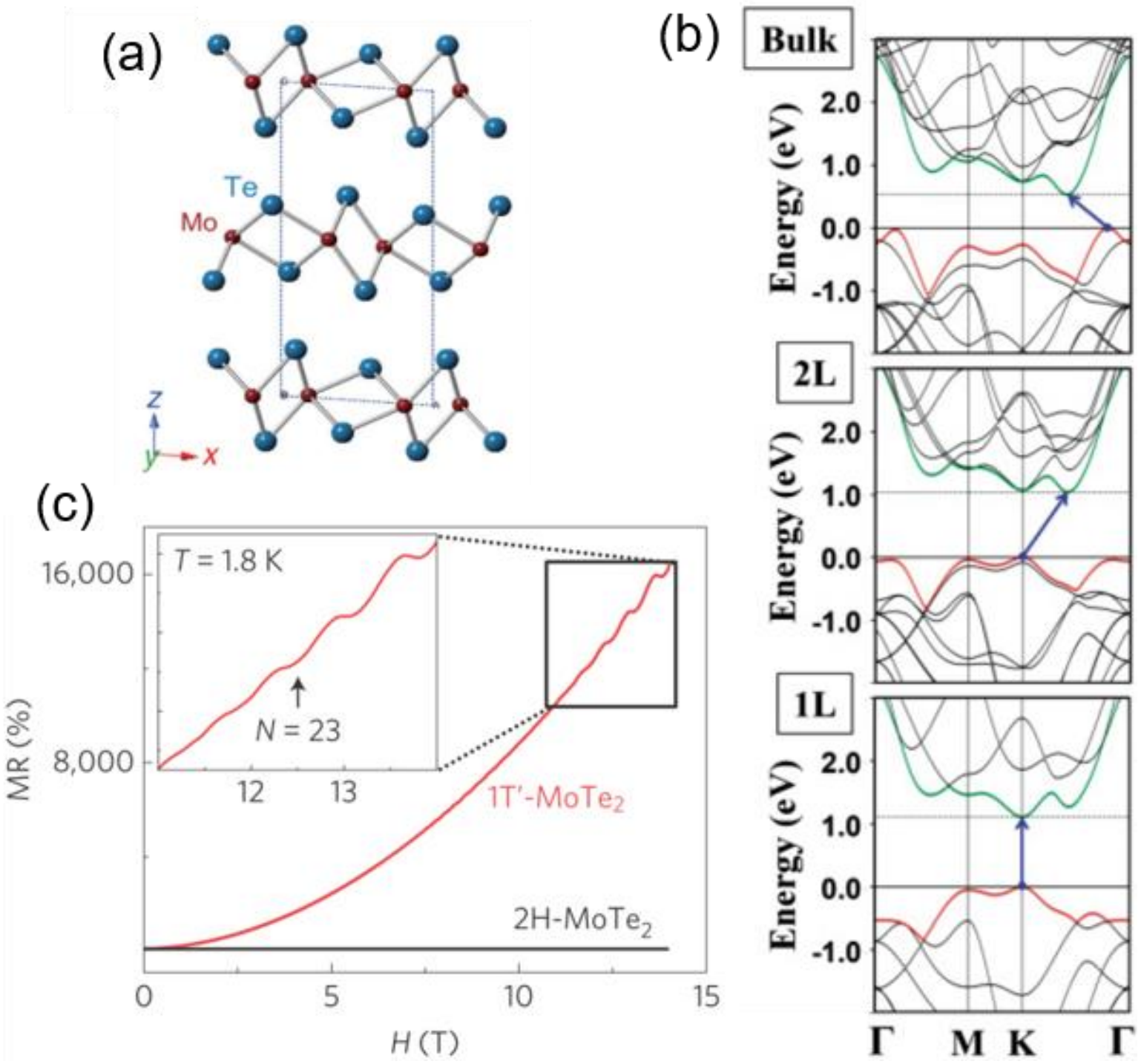

**Figure 23.** (a) Crystal structure of 1T'-$MoTe_2$. Adapted from Ref. [147]. (b) Band structures of 2H-$MoTe_2$. Adapted from Ref. [146]. (c) MR of 1T'- and 2H-$MoTe_2$ taken at 1.8 K. Inset: SdH oscillations of 1T'-$MoTe_2$. Adapted from Ref. [18].

### *4.2 Topological materials*

Topological insulators are the first extensively studied topological materials. The body state is an insulator, and the surface state is protected by topology and presents a metallic state. From the energy band point of view, the body state has a band gap, while the surface state, shaped by the SOC, has a linear band. Electrons can be transported in two conductive channels on the surface of the material, and the direction of motion and the direction of spin are locked, i.e., spin momentum locking. Owning to this feature, the transport process of the surface state is not scattered. Figure 24(a) shows the calculated bulk conduction band and bulk valance band dispersions along high-symmetry directions of the surface Brillouin zone of $Bi_2Te_3$, which is confirmed by the ARPES measurement (figure 24(b)) [148]. It can be seen that the surface state is composed of a nondegenerate Dirac cone and the bulk state has a band gap. Wang et al. observed an MR of 600% in five-layer $Bi_2Te_3$ at 340 K and 13 T (figure 24(c)) [87], which is related to the surface state that gives rise to massless carriers and large mobility (~ $1.02 \times 10^4$ $cm^2$/V s). The MR has a nonmonotonic dependence on temperature.

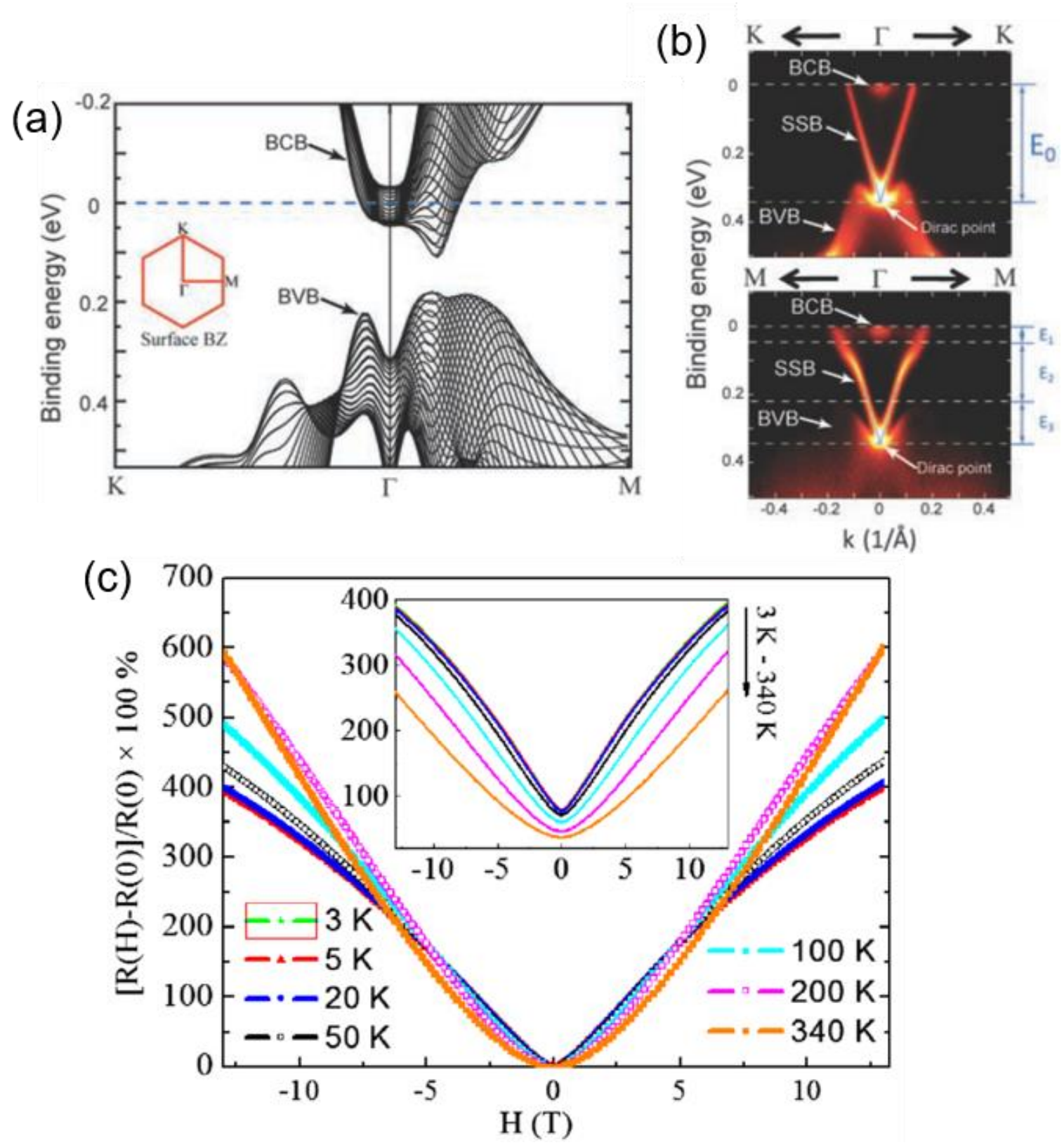

**Figure 24.** (a) Calculated bulk conduction band and bulk valance band dispersions along high-symmetry directions for the surface Brillouin zone (inset). (b) ARPES measurement of band dispersions along K-G-K (top) and M-G-M (bottom) directions. The broad bulk conduction band and bulk valance

band dispersions are similar to those in (a), whereas the sharp V-shape dispersion is from the surface state band. The apex of the V-shape dispersion is the Dirac point. Adapted from Ref. [148]. (c) Field dependence of transverse resistance (inset) and MR ratio between 3 K and 340 K for a $Bi_2Te_3$ nanosheet. Adapted from Ref. [87].

Another large group of topological materials are topological semimetals, which are characterized by symmetry protected band crossings at or near the Fermi level in the Brillouin zone. According to the degeneracy and momentum space distribution of the nodal points, topological semimetals can be divided into three categories, i.e., Dirac semimetals, Weyl semimetals (type I and II) and nodal-line semimetals. The band crossings of 3D Dirac semimetals are isolated points with fourfold degeneracy, called Dirac nodes. When the symmetry of time reversal or space inversion is broken, the fourfold degeneracy becomes twofold, and the Dirac semimetal is transformed into a Weyl semimetal. For Weyl semimetals that lack inversion, they are further classified into type I and type II according to whether there are inclined Weyl nodes. In nodal-line semimetals, the band crossing occurs along a closed curve in the momentum space. The nodal line can take the form of an extended line, a closed loop or a chain. Generally speaking, topological semimetals have intriguing characteristics such as quantum oscillations, nontrivial Berry phase, high carrier mobility, negative LMR and extraordinary XMR.

As a representative Dirac semimetal, $Cd_3As_2$ has a pair of 3D Dirac points in the body and a nontrivial Fermi arc on the surface [149], which is supported by the ARPES experimental results (figure 25(a)) [150]. In addition, a negative LMR caused by chiral anomaly is observed in $Cd_3As_2$ nanowires, as shown in figure 25(b) [151]. SdH quantum oscillation data suggest that there is a nontrivial Berry phase in $Cd_3As_2$ [83]. Cheng et al. have grown $Cd_3As_2$ of different thicknesses by molecular beam epitaxy. From a bulk material to a 50 nm thick nanosheet, the transition from semimetal to semiconductor occurs [152]. Moreover, XMR has been observed by different groups. Feng et al. observed a large linear MR of 3100% at 2 K and 14 T in $Cd_3As_2$ crystals (figure 25(c)) [82]. For $Cd_3As_2$ nanoplates, the XMR reaches 2000% at 200 K and 14 T (figure 25(d)) [126].

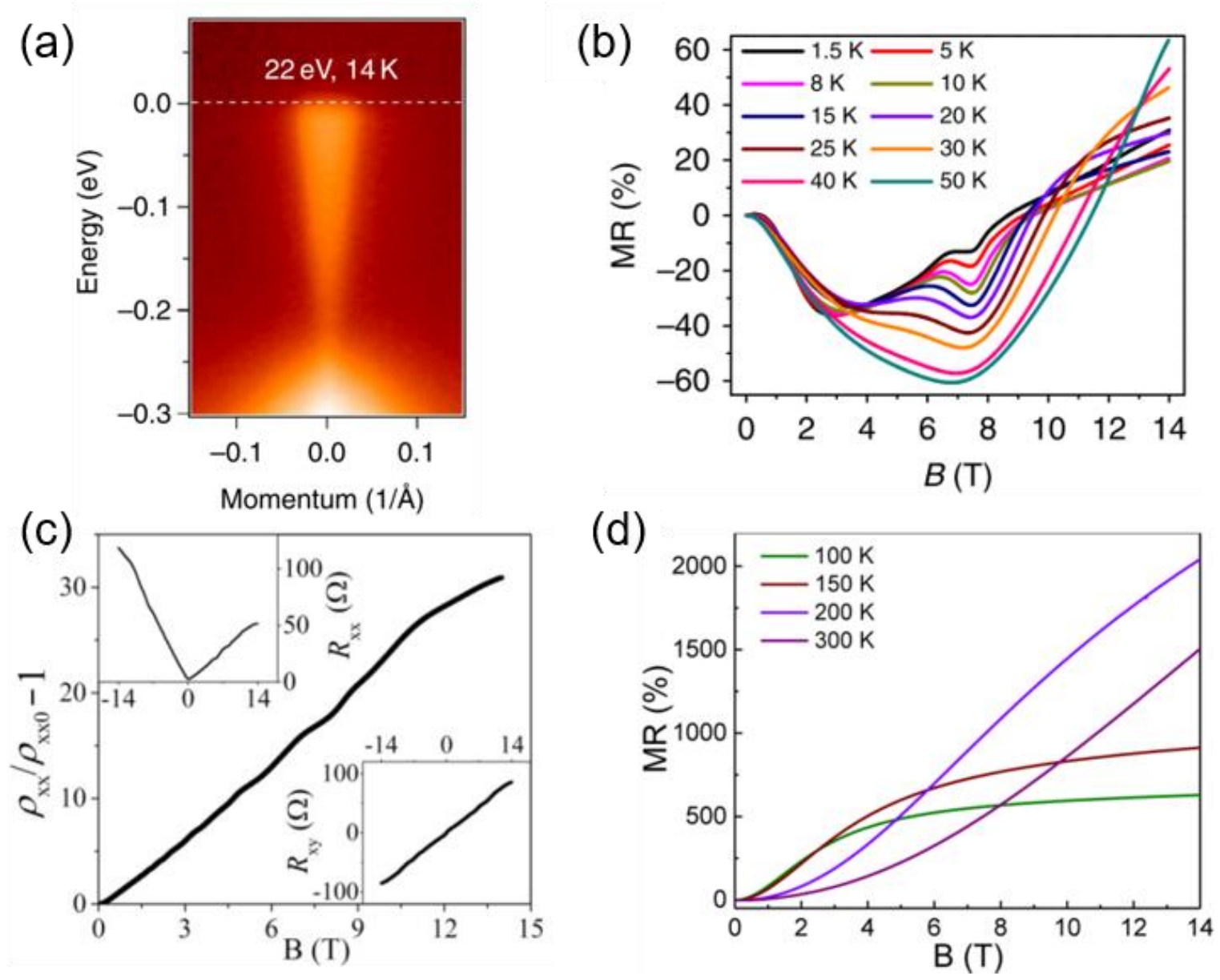

**Figure 25.** (a) ARPES dispersion cut of $Cd_3As_2$ near the Fermi level around the center of the surface Brillouin zone. Adapted from Ref. [150]. (b) Negative LMR measured at temperatures from 1.5 K to 300 K [151]. (c) MR of $Cd_3As_2$ crystal taken at 2 K. The data have been symmetrized by averaging over the positive and negative field directions. Upper inset: original data $R_{xx}$. Lower inset: Hall data $R_{xy}$. Adapted from Ref. [82]. (d) MR measured at different temperatures for $Cd_3As_2$ nanoplates. Adapted from Ref. [126].

ZrSiS is a potential candidate for nodal-line semimetals. By the ARPES measurement, it is suggested that ZrSiS has an electronic band structure that hosts several Dirac cones, forming a Fermi surface with a diamond-shaped line of Dirac nodes. The energy range of the linearly dispersed bands is as high as 2 eV above and below the Fermi level, which is much larger than other known Dirac materials [153]. Figure 26(a) shows the negative LMR in the transport measurement [115], which is supposed to relate to the Adler-Bell-Jackiw chiral anomaly of 3D Dirac fermions in the material. The XMR of ZrSiS reaches up to 1.4 $\times 10^5$% at 2 K and 9 T with current along the *a* axis and magnetic field parallel to the *c* axis (figure 26(b)). Such a prominent XMR should originate from the robust existence of multiple Dirac nodes in the nodal-line semimetal.

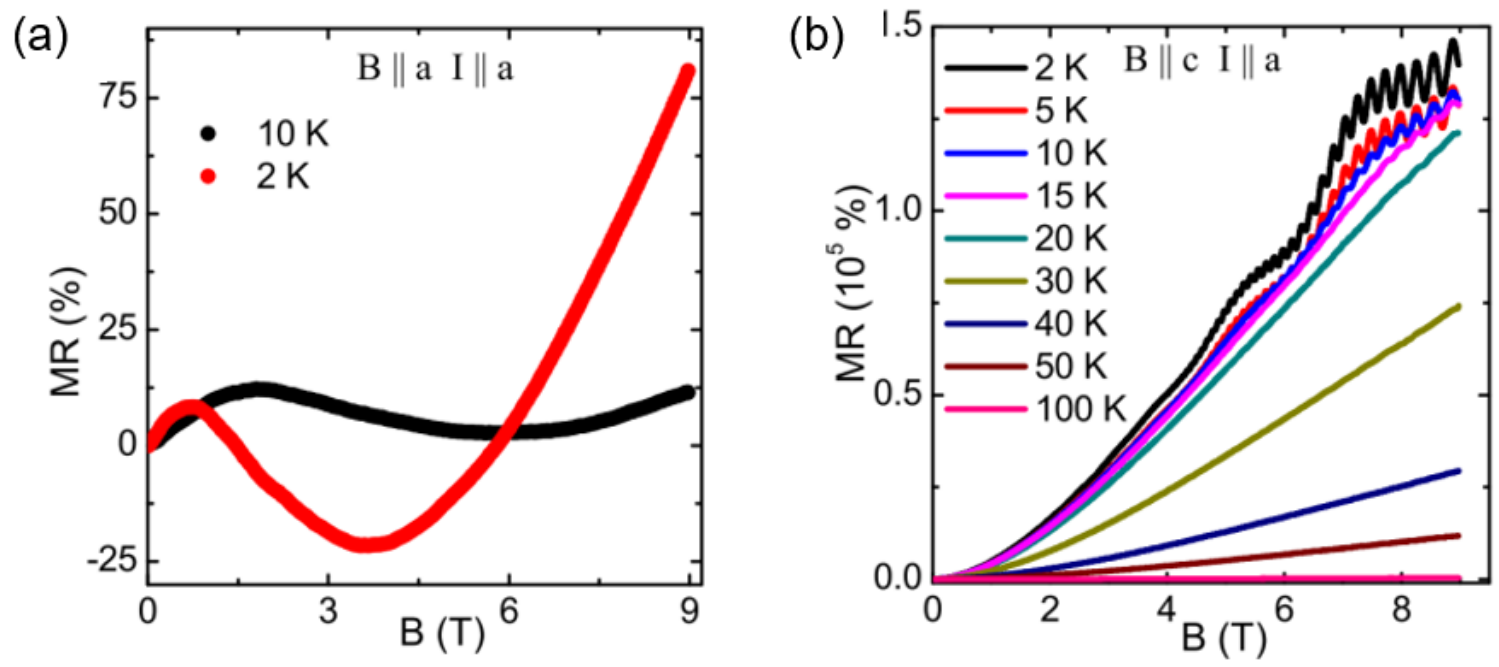

**Figure 26.** (a) LMR with current and field along the *a* axis. (b) Transverse MR with current along the *a* axis and magnetic field parallel to the *c* axis. Adapted from Ref. [115].

As mentioned above, several monopnictides were first experimentally confirmed Weyl semimetals. Weng et al. theoretically predicted that TaAs is a Weyl semimetal which lacks inversion symmetry but maintains time reversal symmetry [97]. It followed that Lv et al. observed Fermi arcs and paired Weyl cones with opposite chirality at and near the Weyl points through ARPES, as shown in figure 27(a) and (b). [154, 155]. These are direct evidence that TaAs is a Weyl semimetal. In the transport measurements, an obvious negative LMR effect appeared and was linked to chiral anomaly (figure 27(c)) [12]. In addition, an XMR as high as $8 \times 10^4$% was observed at 1.8 K and 9 T (figure 27(d)), with an ultrahigh electron mobility of up to $1.8 \times 10^5$ $cm^2/V$ s. According to the discussion in the mechanism section, the electron-hole compensation and high mobility associated with Weyl fermions should contribute to XMR.

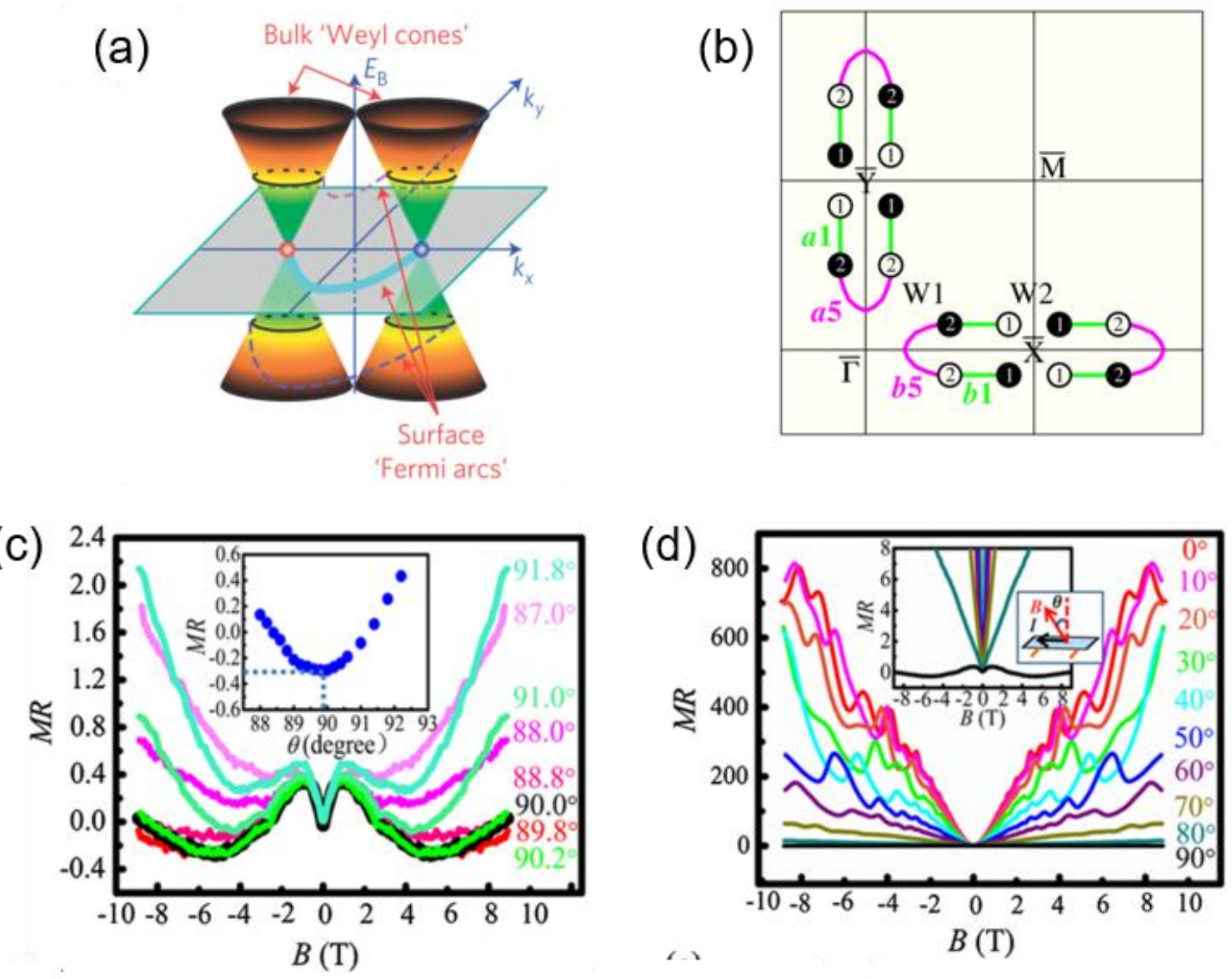

**Figure 27.** (a) Surface projection of a pair of bulk Weyl cones with opposite chirality connected by a surface Fermi arc. Blue, cyan and magenta curves are Fermi arcs at different binding energies. Adapted from Ref.[155]. (b) Schematic of the Fermi-arc connecting pattern. Solid and hollow circles represent the projected Weyl nodes with opposite chirality, and numbers inside indicate the total chiral charge. Adapted from Ref. [154]. (c) LMR measured around $\theta = 90^{o}$, showing obvious negative LMR. $\theta = 90^{o}$ indicates $B$ // $I$. Inset: Minima of LMR curves at different angles in a magnetic field from 1 to 6 T. (d) Angular- and field-dependence of MR in a TaAs single crystal taken at 1.8 K. Adapted from Ref. [12].

Unlike TaAs (classified as a type I Weyl semimetal), type II Weyl semimetals possess Weyl points that appear at the boundary of the electron and hole pockets, resulting in highly inclined Weyl cones. Alexey et al. predicted that $WTe_2$ is a type II Weyl semimetal in theory [156]. And experimentally, Wang et al. observed the electronic structure of $WTe_2$ through ARPES, which is in good agreement with theoretical calculations, as shown in figures 28(a) and 28(b) [157]. The observation of negative LMR induced by chiral anomaly in $WTe_2$ further proves its Weyl semimetal nature (figure 28(c)) [145]. The XMR of $WTe_2$ has been reviewed above, and its mechanism has been discussed in detail in the mechanism section.

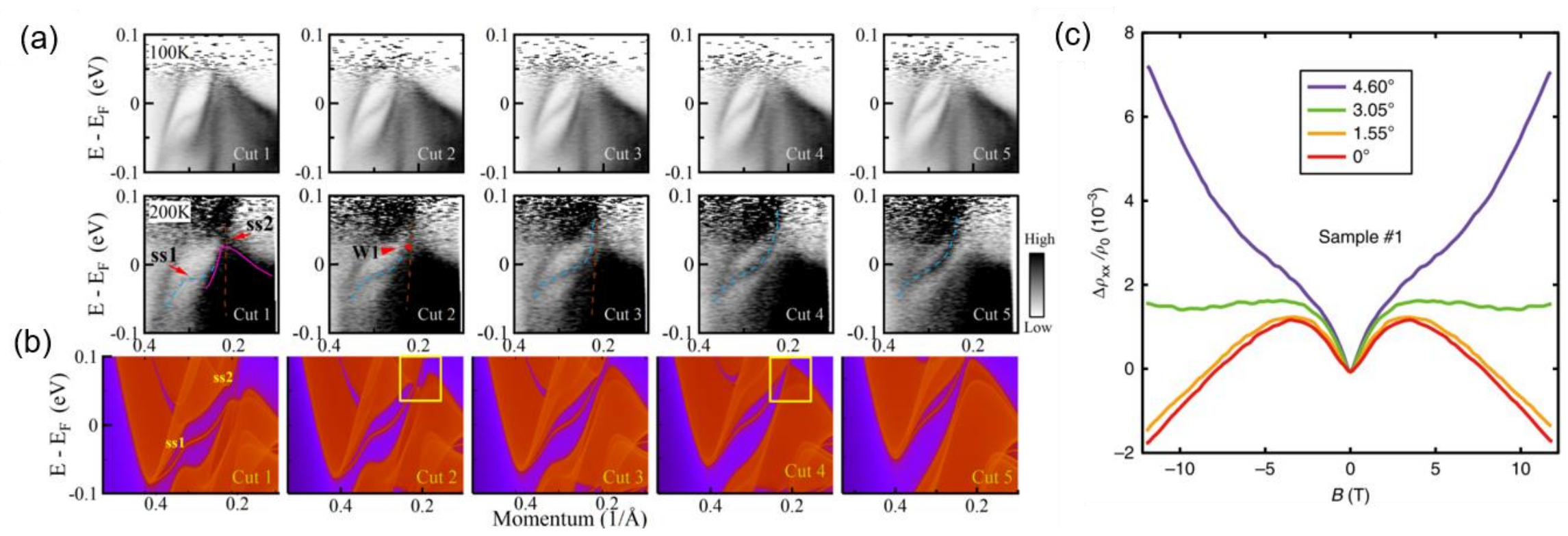

**Figure 28.** (a) Band structures of $WTe_2$ measured along different momentum cuts at 100 and 200 K. (b) Calculated band structures along the same momentum cuts. Adapted from Ref. [157]. (c) Sample exhibits a negative LMR only in a narrow angle region around *B* // *I*. Adapted from Ref. [145].

$Co_3Sn_2S_2$ represents another group of Weyl semimetals, which lack time reversal symmetry but maintain inversion symmetry. Through ARPES measurements, the electronic structure of $Co_3Sn_2S_2$ was characterized. The Fermi arc on the surface and the linear bulk band dispersion were found, as shown in figures 29(a) and 29(b) [158]. Morali et al. determined the Fermi arc connectivity between the Weyl points in Brillouin zone by scanning tunneling spectroscopy [159]. Furthermore, the transport measurements revealed a negative LMR (equivalent to a positive longitudinal magnetoconductance) caused by chiral anomaly (figure 29(c)), as well as a nontrivial Berry phase retrieved from SdH quantum oscillation data [160]. However, the MR of $Co_3Sn_2S_2$ recorded as 220% at 3 K and 9 T is not outstanding (figure 29(d)) [26]. This may be due to the existence of magnetic ordering and magnetic scattering. From the magnitude of MR, $Co_3Sn_2S_2$ is not a typical XMR material. In order to completely describe the MR properties of topological semimetals, it is still included here.

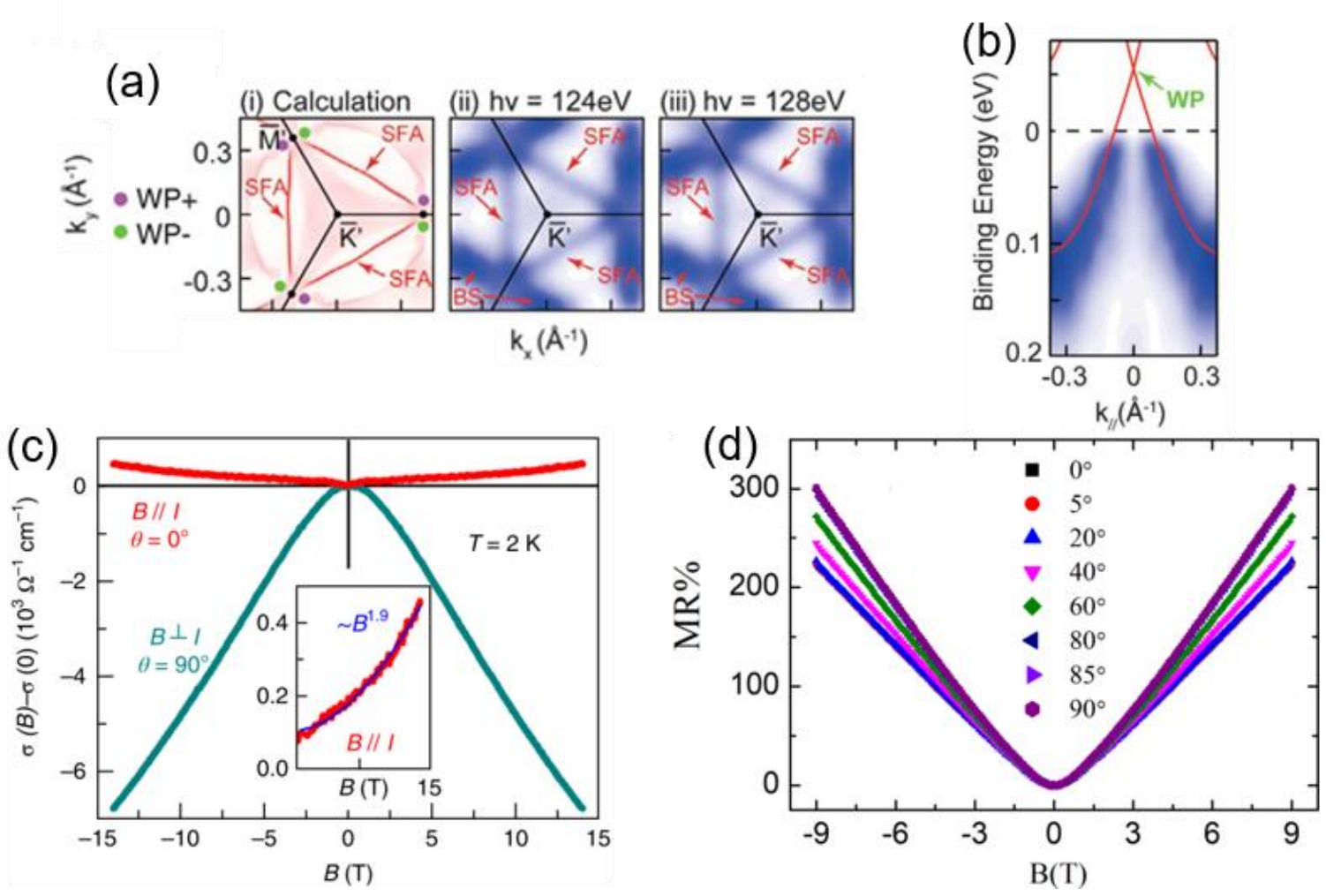

**Figure 29.** (a) Comparison of (i) the calculated Fermi surface from both bulk and surface states and (ii)(iii) the experimental Fermi surfaces under different photon energies. (b) Band dispersion showing linear dispersions toward the Weyl point above the Fermi level, in agreement with the calculations (red curves). Adapted from Ref. [158]. (c) Magnetoconductance at 2 K in both cases of $B \perp I$ and $B // I$. Adapted from Ref. [160]. (d) Angle dependent MR. Adapted from Ref. [26]

## 5. Conclusions and outlook

To conclude, in this paper we have reviewed the history of MR research, various MR systems and main mechanisms. The current focus is on XMR materials in metals or semimetals and their unique properties that are highly correlated with XMR. The electron-hole compensation is considered to be the main driving force of XMR, and high carrier mobility exists as a promotive factor in most XMR materials. In addition, many XMR materials exhibit novel and interesting characteristics, such as a layered structure and topological quasiparticles. The representative 2D materials and topological materials with a significant XMR have been further reviewed, with an emphasis on their relationship with XMR. Despite the massive discovery of XMR materials and the in-depth investigation of the origin, there are still some issues that need to be resolved.

(i) A material with simple structure, environmentally friendly chemical composition and outstanding XMR is the central task of XMR research. According to the current research results, the design and development of new materials to obtain better performance is still an important issue.

(ii) The mechanisms are still under debate. A universal law does not exist. For some XMR materials, an unambiguous dominant mechanism may apply. However, for the rest numerous cases, various mechanisms have been proposed. Microscopic pictures of these mechanisms should be checked through elaborately designed experiments. The discovery of a new mechanism is of special interest and importance for physical understanding.

(iii) Owning to the overlap of XMR, 2D and topological materials, emerging properties are expected to generate new interdisciplinary applications. The possible scenarios need to be expanded. 2D materials have inherent advantages in the construction of miniaturized devices. Therefore, it is an important direction to utilize the layered feature and XMR property of 2D XMR materials.

Relatedly, it has been found that the XMR of layered materials depends on the thickness or number of layers of the material, such as in $WTe_2$ and $Cd_3As_2$ nanosheets. The reasons need to be further explored.

(iv) As we know, factors such as high mobility, steep band and Dirac fermions can all promote XMR. Why is there no outstanding MR in graphene (~ 200%) which hosts all of the features?

(v) Although the practical application research of XMR materials is still at a very preliminary stage, some issues still need to be raised for future research. The huge external field and cryogenic environment are two aspects required to obtain XMR, which hinder practical application. XMR working at room temperature without high external field has always been the ultimate goal of MR research. In addition, regarding consumption and integration problems, a lot of research and engineering work will be needed to design and manufacture specific devices. The work may include prototype design, fabrication quality, driving mode, packaging technology, high-density integration, etc. Finally, air stability is another key factor of future research.

Anyway, XMR is a fundamental, important and fruitful research direction in modern condensed matter physics, for both theoretical and experimental researchers. Future research work calls for more insightful perspectives, especially in interdisciplinary fields.

## Acknowledgments

We appreciate discussions with W. L. Zhen and W. Tong and support from the National Natural Science Foundation of China under Grant Nos. 11874363 and U1932216.